\documentclass[smallextended]{svjour3}
\renewcommand{\makeheadbox}{}
\usepackage{fix-cm}
\usepackage{natbib}

\usepackage[utf8]{inputenc}
\usepackage{longtable}
\usepackage{pdflscape}
\usepackage{graphicx}
\usepackage{multirow}
\usepackage{amsmath,amssymb,amsfonts}
\usepackage{mathrsfs}
\usepackage[title]{appendix}
\usepackage{textcomp}
\usepackage{manyfoot}
\usepackage{booktabs}
\usepackage{algorithm}%
\usepackage{algorithmicx}%
\usepackage{algpseudocode}%
\usepackage[most]{tcolorbox}
\usepackage{enumitem}
\usepackage{pifont}
\usepackage[hidelinks]{hyperref}
\usepackage{tabularx}
\usepackage{xcolor}
\usepackage{verbatim}
\usepackage{listing}
\usepackage{xurl}
\definecolor{toolgray}{gray}{0.4}

\tcbset{
  resbox/.style={
    colback=gray!5,
    colframe=black,
    boxrule=0.5pt,
    arc=1pt,
    left=3pt,
    right=3pt,
    top=3pt,
    bottom=3pt
  }
}

\usepackage{listings}
\usepackage{xcolor}

\definecolor{codeblue}{RGB}{40,80,140}
\definecolor{codegreen}{RGB}{70,120,70}
\definecolor{codegray}{RGB}{110,110,110}

\lstdefinelanguage{yaml}{
  keywords={name,on,push,pull_request,jobs,runs-on,steps,uses,with,languages,build-mode},
  keywordstyle=\color{codeblue}\bfseries,
  comment=[l]{\#},
  commentstyle=\color{codegreen}\itshape,
  stringstyle=\color{codegray},
  sensitive=true
}

\usepackage{tikz}
\usepackage{graphicx} 
\usetikzlibrary{arrows.meta,positioning,fit,backgrounds,calc}
\usepackage{longtable}

\newcommand{\code}[1]{\textsf{\small{}#1}}
\newcommand{\codesm}[1]{\textsf{\scriptsize{}#1}}

\newlist{onlineres}{enumerate}{1}
\setlist[onlineres]{%
  label={\scriptsize[\arabic*]}, 
  ref=\arabic*, 
  nosep,
  font=\scriptsize,
  before=\scriptsize
}

\makeatletter
\newcommand{\olcite}[1]{%
  \textsuperscript{[%
    \@tempcnta=0
    \@for\@key:=#1\do{\advance\@tempcnta by 1}%
    \@tempcntb=0
    \@for\@key:=#1\do{%
      \advance\@tempcntb by 1
      \hyperref[ol:\@key]{\scriptsize\ref{ol:\@key}}%
      \ifnum\@tempcntb<\@tempcnta ,\fi
    }%
  ]}%
}
\makeatother

\newcounter{reporow}

\makeatletter
\newcommand{\repocite}[1]{%
  \textsuperscript{[%
    \@tempcnta=0
    \@for\@key:=#1\do{\advance\@tempcnta by 1}%
    \@tempcntb=0
    \@for\@key:=#1\do{%
      \advance\@tempcntb by 1
      \hyperref[repo:\@key]{\scriptsize\ref*{repo:\@key}}%
      \ifnum\@tempcntb<\@tempcnta ,\fi
    }%
  ]}%
}
\makeatother

\begin{document}

\title{Characterizing GitHub Security Actions Alerts For Web Application Tutorials}


\titlerunning{Characterizing Github Security Action Alerts}        

\author{Bhagya Chembakottu        \and
        Martin P. Robillard 
}


\institute{McGill University \at
              Montreal, Canada \\
              \email{bhagya.chembakottu@mail.mcgill.ca}           
           \and
           McGill University \at
              Montreal, Canada \\
              \email{robillard@acm.org}
}

\date{}

\maketitle

\begin{abstract}
The high frequency of cybersecurity incidents involving web applications highlights the importance of secure software development practices. Static Application Security Testing (SAST) tools help identify potential vulnerabilities in source code without executing the code. Integrating SAST tools into GitHub Security Actions enables security alerts to be reported during development. However, the characteristics of these alerts can make them difficult for developers with limited security expertise to understand, prioritize, and address. Many tutorials for learning web application development also involve source code with security implications. We investigated the properties of SAST alerts relevant to this learning context. We analyzed 6,609 alerts from four default GitHub Security Actions configurations across 60 Django and Spring tutorial repositories. We evaluated the alerts along four dimensions: alert volume, alert relevance, conceptual breadth, and remediation usability. Of the 6,609 alerts, 3,750 (56.7\%) targeted tutorial-related files, while 2,859 (43.3\%) targeted other repository files. The configurations triggered between 34 and 646 distinct rules, and the five most frequently triggered rules accounted for 25.2\%--68.2\% of alerts, depending on the configuration. Across configurations, 0.4\%--32.1\% of alert messages named concrete framework or code constructs, and 0\%--8.1\% met our operational criteria for remediation specificity. Overall, default GitHub Security Actions report a diverse set of security alerts, but these alerts do not consistently focus on tutorial code or provide concrete guidance for addressing the reported issues.

\end{abstract}

\section{Introduction}
\label{sec:introduction}

GitHub reported a 172\% year-over-year increase in vulnerabilities related to Broken Access Control in repositories, the highest-ranked security risk in the Open Worldwide Application Security Project (OWASP) \citep{owasp_top_ten}. This increase indicates a need for security practices in the early stages of development. One potential mitigation involves leveraging GitHub Actions. GitHub describes Actions as a way to ``automate, customize, and execute software development workflows right in your repository'' \citep{github_actions_docs}. GitHub reports 11.5 billion minutes of total Actions usage in 2025 \citep{octoverse2025}. We use \emph{GitHub Security Actions} to refer to workflows that run static security analysis through GitHub Actions and report alerts through repository workflows \citep{gitapi}. Developers can configure the analysis to use tools such as CodeQL \citep{codeql1}, SonarCloud \citep{sonarcloud1}, and Semgrep \citep{semgrep1}.


Access to automated analysis, however, does not guarantee that developers can interpret or act on its output. Security tools have limited effect when developers do not adopt them \citep{barriers}, do not understand their output \citep{suppressed_warning}, or cannot connect alerts to concrete development decisions \citep{context_matter}. \citet{action_required2026} found adoption rates from 0.6\% to 52.9\% for GitHub Actions security practices in the repositories they studied. They linked this variation to low awareness, confusion about when the practices apply, and limited setup knowledge. \citet{didario2023securitytesting} and \citet{charoenwet2024secure} found differences in how companies conduct security testing and a lack of standardized guidelines for security testing. Prior work therefore shows that access to security automation alone does not determine how developers use or interpret its alerts.


Many developers learn web application development through online tutorials \citep{arya2023learningresources,salerno2026toolinstallation}, and these tutorials and code examples may contain insecure practices that can propagate insecure code patterns to other projects \citep{chen2024codereuse}. Tutorials can provide code repositories, or developers may create repositories while following the material. Tutorial authors may simplify code to teach one concept at a time. Security beginners may not know which simplifications apply only to the tutorial context or require changes before production use. By \emph{security beginner}, we refer to a developer who can follow a programming tutorial but may not yet recognize the security implications of framework-related features.

Tutorial repositories can be widely visible. For example, Baeldung's \code{eugenp/tutorials} repository had approximately 37.3K stars and 53.2K forks as of September 30, 2026 \citep{github_baeldung_tutorials}. Although these figures do not establish how developers use the repository, they illustrate the visibility of tutorial code. Because tutorials focus on functional requirements while security remains an important non-functional requirement \citep{yu2023nfrtesting}, developers can use GitHub Security Actions to check tutorial code for security issues. The security tools accessible through GitHub Actions report alerts with severity levels, rule descriptions, and sometimes suggested fixes \citep{comparison}. This format supports vulnerability triage and mitigation by experts, but it does not necessarily support beginner-oriented learning. For this reason, we ask:

\begin{quote}
\textit{To what extent are alerts from default GitHub Security Actions configurations applicable to tutorials?}
\end{quote}

We study this question through four quantitative dimensions:

\noindent\textbf{Alert volume:}
How many alerts does a developer see when running GitHub Security Actions with default configurations on a tutorial repository?

\noindent\textbf{Alert relevance:}
Do alerts target the code that the tutorial focuses on, or do they target dependencies, build artifacts, workflow files, etc.?

\noindent\textbf{Conceptual breadth:}
How many distinct security rules and concepts do the tools cover, and do alerts concentrate around recurring security concepts?

\noindent\textbf{Remediation usability:}
Do alert messages name concrete framework or code constructs and provide specific remediation steps that a developer can connect to Django or Spring security practices?

We analyzed GitHub Security Actions output across 60 web application tutorial repositories: 37 Django repositories and 23 Spring repositories. We reviewed 350 CodeQL alerts, 4,206 Semgrep alerts, 1,055 SonarCloud alerts, and 998 SonarCloud Security Hotspots. We assessed whether each alert targeted tutorial-context code, which security concepts each alert covered, whether each message referred to concrete framework constructs, and whether each message provided specific remediation guidance. We also mapped the triggered rules to the Common Weakness Enumeration (CWE) and OWASP Top 10 \citep{owasp_top_ten}.

Our results show differences between the four default configurations. First, 3,750 of the 6,609 alerts (56.7\%) targeted tutorial-related files, while 2,859 (43.3\%) targeted other repository files. Of the non-tutorial alerts, 2,331 targeted dependency manifests or build metadata and 422 targeted GitHub Actions workflow files. The proportion of tutorial-scoped alerts ranged from 45.7\% for Semgrep to 80.6\% for SonarCloud Security Hotspots. Second, the configurations triggered between 34 distinct rules (Hotspots) and 646 (Semgrep). The five most frequent rules accounted for 25.2\%--68.2\% of alerts, depending on the configuration. Third, concrete framework or code references appeared in 0.4\%--32.1\% of alert messages, while 0\%--8.1\% met our operational criteria for remediation specificity. Finally, alert volume varied across configurations and repositories. CodeQL reported alerts in 41 of the 60 repositories, while Hotspots reported alerts in 58. A single Semgrep scan of GH-SB-FullStack reported 925 alerts. Repository structure and tool choice are associated with differences in the amount and type of security alerts that developers receive.

We conclude that the default SAST configurations do not provide a single, consistent view of security in tutorial repositories. Tool choice changes the alert volume, repository scope, rule distribution, and remediation information that developers see. Repository structure can also change alert volume. Based on our observations, we propose guidelines for adapting SAST output to tutorial code. This paper makes the following contributions:

\begin{enumerate}
    \item An evaluation framework for SAST alerts from tutorial repositories. We evaluate alerts across four dimensions: alert volume, alert relevance, conceptual breadth, and remediation usability.

    \item A comparative empirical analysis of 6,609 alerts from four default GitHub Security Actions configurations across 60 Django and Spring tutorial repositories.

    \item A public research artifact that includes a manually curated corpus of 60 Django and Spring tutorial repositories, GitHub Actions workflows, raw alert data, our alert annotations, security-concept mappings, and statistical analysis code. The artifact supports replication and future comparative studies.
\end{enumerate}

The remainder of this paper is structured as follows. Section 2 details a motivating example from our dataset to explain the problem statement. Section 3 discusses relevant past research. Section 4 describes the study design, data collection choices, and data analysis. Section 5 presents the results of the evaluation study, followed by a discussion and conclusion in Sections 6 and 7. This article is complemented by an online dataset \citep{psp_replication_package}.

\section{Motivating Example}
\label{sec:motivating-example}

In February 2022, GitHub introduced GitHub Security Actions workflow templates, allowing developers to add a SAST tool of their choice with a few clicks from the Actions tab of each repository rather than manually adding a workflow under \code{.github/workflows/}~\citep{github_security_changelog_2022,github_security_workflows_2022}. Figure ~\ref{fig:motivating-example} shows a CodeQL workflow template that can be added to a Python repository~\citep{github_actions_docs,codeql1} by simply selecting New Workflow $\rightarrow$ CodeQL $\rightarrow$ Configure $\rightarrow$ Commit.
We consider \texttt{GH-DJ-Allauth} (\texttt{django-allauth}), one of the Django tutorial repositories in our study. In our collected data, CodeQL reported 27 alerts, Sonar reported 35 alerts, hotspot reported 31 alerts reported and Semgrep reported 301 alerts for this repository.

\begin{figure}[!htbp]
  \centering
  \includegraphics[
    width=\linewidth,
    trim={0 0 3cm 0},
    clip
  ]{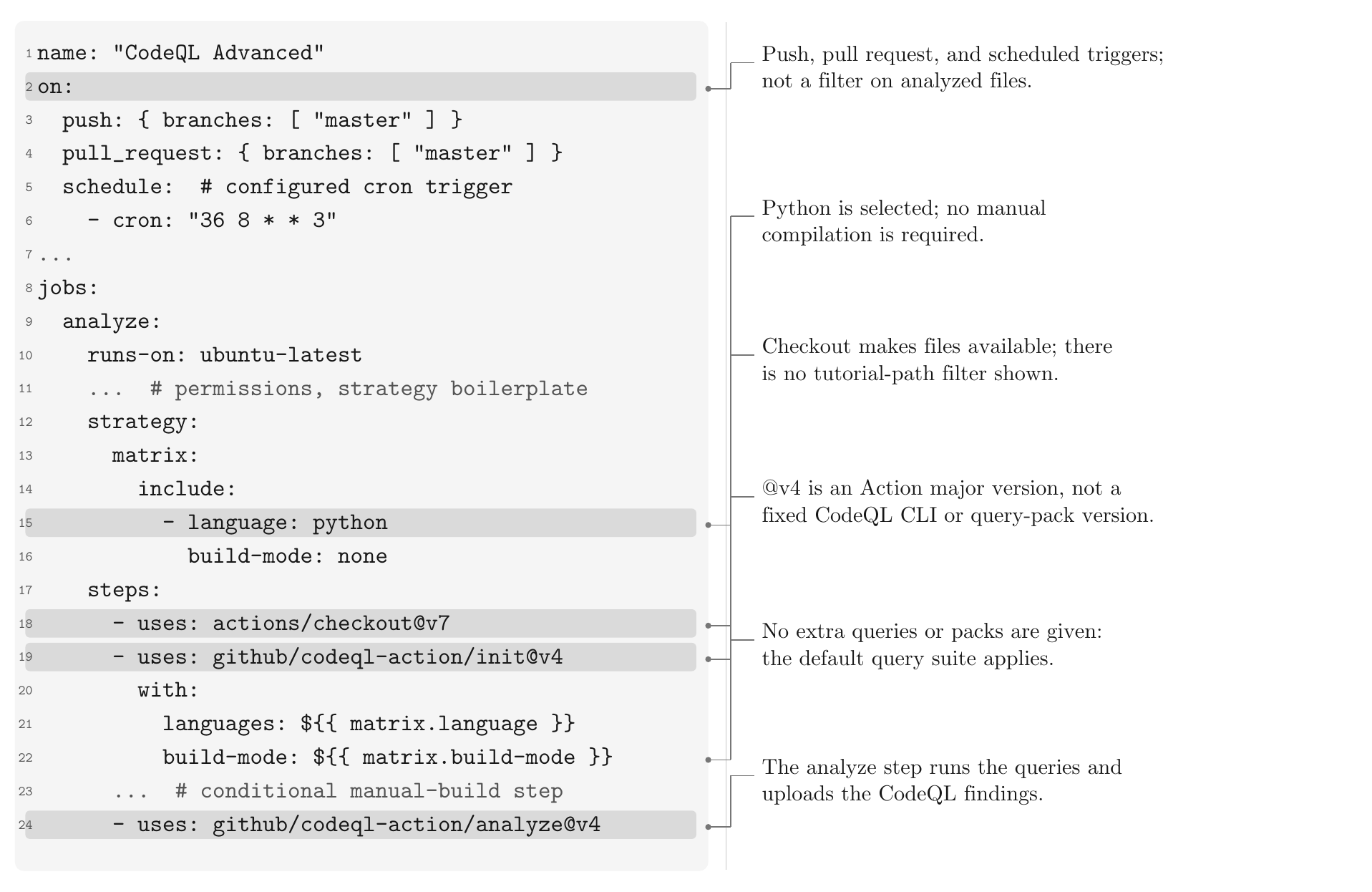}
  \caption{Selected settings of the CodeQL workflow template for Python (14 October 2026). The image omits default actions related boilerplate and the conditional manual-build block. The annotations distinguish when the workflow runs, the language and files analyzed, and the queries that produce the reported alerts.}
  \label{fig:codeql-workflow}
\end{figure}

\begin{figure}[!htbp]
\centering
\resizebox{0.69\linewidth}{!}{%
\begin{tikzpicture}[
    x=1mm,y=1mm,
    every node/.style={anchor=north west,inner sep=0pt,outer sep=0pt,
      align=left,font=\sffamily\fontsize{8.3}{10}\selectfont},
    rule/.style={draw=black!70,line width=0.4pt},
    flow/.style={draw=black!80,line width=0.6pt,
      -{Stealth[length=1.4mm,width=1mm]}},
    code/.style={font=\ttfamily\fontsize{6.3}{8}\selectfont},
    dimension/.style={text width=25mm,font=\sffamily\bfseries\fontsize{8.3}{10}\selectfont},
    outputA/.style={text width=24mm},
    outputB/.style={text width=28mm}
]
\node[anchor=north] at (42,0) {\textbf{GH-DJ-Allauth}\quad Django request view};
\draw[rule] (0,-6) rectangle (84,-43);
\node at (2,-8) {\textbf{V\enspace\code{allauth/idp/oidc/views.py}}};
\node[anchor=north east] at (82,-8) {Reported by \textbf{A + B}};
\node[code] at (2,-13) {
   162\enspace signed\_request\_info = request.POST.get("request")\\
   \phantom{162}\enspace \ldots\\
   176\enspace \ldots{} signer.unsign\_object(signed\_request\_info)
};
\draw[rule] (2,-26) -- (82,-26);
\node at (2,-28) {\textbf{D\enspace\code{frontend/Dockerfile}}};
\node[anchor=north east] at (82,-28) {Reported by \textbf{B only}};
\node[code] at (2,-33) {
    1\enspace FROM node:18\\
    \phantom{1}\enspace \ldots\\
    10\enspace CMD ["npm", "start"]
};
\draw[rule,line width=0.6pt] (55,-43) -- (55,-46);
\draw[flow] (55,-46) -| (40,-49);
\draw[flow] (55,-46) -| (70,-49);
\node at (0,-50) {Selected alerts};
\node[anchor=north] at (40,-50) {\textbf{Config. A: CodeQL}};
\node[anchor=north] at (70,-50) {\textbf{Config. B: Semgrep}};
\draw[rule,line width=0.6pt] (0,-55) -- (84,-55);
\node[dimension] at (0,-57) {Alert volume};
\node[outputA] at (28,-57) {V (27 alerts total)};
\node[outputB] at (56,-57) {V + D (301 total)};
\draw[rule] (0,-63) -- (84,-63);
\node[dimension] at (0,-65.3) {Alert relevance};
\node[outputA] at (28,-65.3) {V: Django view};
\node[outputB] at (56,-65.3) {V: Django view\\D: Dockerfile};
\draw[rule] (0,-74.5) -- (84,-74.5);
\node[dimension] at (0,-77) {Conceptual\\breadth};
\node[outputA] at (28,-77) {V: CWE-601\\5 distinct rules};
\node[outputB] at (56,-77) {V: CWE-601\\D: CWE-250\\114 distinct rules};
\draw[rule] (0,-90) -- (84,-90);
\node[dimension] at (0,-92.5) {Remediation\\usability};
\node[outputA] at (28,-92.5) {V: issue named};
\node[outputB] at (56,-92.5) {V: validate URL\\D: non-root USER};
\draw[rule,line width=0.6pt] (0,-105) -- (84,-105);
\end{tikzpicture}%
}
\caption{Selected code excerpts and reported security alerts from \code{GH-DJ-Allauth}. A refers to CodeQL and B to Semgrep, with both reporting the view alerts (V) and B also reporting the Dockerfile alerts (D). The counts in parentheses refer to all alerts reported for the repository, not just V and D. Code excerpts are attributed to the \href{https://github.com/research-projects-all/django-allauth/tree/bca85ce9075b7bbeda763a1764720d7abc2b3f38}{archived repository}, reproduced under its \href{https://github.com/research-projects-all/django-allauth/blob/bca85ce9075b7bbeda763a1764720d7abc2b3f38/LICENSE}{MIT license} (Copyright \copyright{} 2010--2026 Raymond Penners and contributors). CWE labels are from the collected alerts; neither a reported alert nor a source excerpt alone establishes a vulnerability.}
\label{fig:motivating-example}
\end{figure}

Figure~\ref{fig:codeql-workflow} shows the relevant parts of the default CodeQL workflow for Python and the parts responsible for default security alert that developer receives. The \code{github/codeql-action/init@v4} step (line~19) initializes CodeQL. The \code{v4} value (line~19) identifies the major version of the GitHub Action, not a fixed version of the CodeQL analysis engine. No additional configuration is specified in the workflow shown (lines~19--22), so CodeQL uses its default set of security checks for Python. This is a predefined subset of the available queries.\footnote{GitHub's \href{https://docs.github.com/en/code-security/concepts/code-scanning/codeql/codeql-query-suites}{CodeQL query suites} and \href{https://docs.github.com/en/code-security/reference/code-scanning/workflow-configuration-options}{workflow configuration options}.} The \code{analyze} step (line~24) runs these security checks and reports the alerts. Developers can customize the analysis by selecting additional security checks or restricting the files analyzed. Figure~\ref{fig:motivating-example} illustrates the alert from the default configurations of CodeQL (Configuration A) and Semgrep (Configuration B). We use alerts in the Django view and the Dockerfile as \textbf{V} and \textbf{D}, respectively. The source lines are selected from the tutorial \texttt{GH-DJ-Allauth} (archived in our dataset).\footnote{Reference: \href{https://github.com/research-projects-all/django-allauth/tree/bca85ce9075b7bbeda763a1764720d7abc2b3f38}{https://github.com/research-projects-all/django-allauth/}} In our data, CodeQL and Semgrep both report a CWE-601 alert in \code{allauth/idp/oidc/views.py} at line 176; Semgrep also reports a CWE-250 alert in \code{examples/react-spa/frontend/Dockerfile}.

The first dimension is \emph{alert volume}. As shown in Figure~\ref{fig:motivating-example}, Configuration A reports alerts on the view (\textbf{V}), whereas Configuration B reports alerts on the view and Dockerfile (\textbf{V} and \textbf{D}). Across the complete repository scan, CodeQL reports 27 alerts and Semgrep reports 301.  The developer can receive different amounts of security alerts from different tools for the same development task.

The second dimension is \emph{alert relevance}. The view alerts \textbf{V} refers to Django code, whereas the additional alerts \textbf{D} targets a Dockerfile. Their relevance depend on their relationship to the tutorial task. A repository can contain application code, tests, deployment files, workflow definitions, examples, dependency specifications, and other supporting artifacts. A SAST tool can report alerts on any of
these artifacts. In the context of a tutorial, a developer may only be able to reason about a subset of them. A technically valid alert can fall outside the concepts that the tutorial asks the developer to inspect or modify.

The third dimension is \emph{conceptual breadth}. Both tools associate \textbf{V} with CWE-601, while Semgrep associates \textbf{D} with CWE-250. Out of all the alerts for this repository, CodeQL triggered five distinct rules and Semgrep triggered 114. Consequently, the configurations differs in the number of rules as well as the alerts that each tool reports.

The fourth dimension is \emph{remediation usability}. For \textbf{V}, CodeQL reports \emph{``Untrusted URL redirection depends on a user-provided value.''} Semgrep describes the possibility of an open redirect and recommends validating the domain against an approved list. Both reported the potential vulnerability but did not provide to a concrete solution, such as Django’s \texttt{redirect()} function.

As this example shows, running default security actions on a tutorial repository can produce different amounts, focus and types of security alert depending on the selected tool. These differences motivate our empirical analysis on the four dimensions across the 60 tutorial repositories.

\section{Related Work}
\label{sec:related-work}

Static application security testing can be a simple and effective way of identifying vulnerabilities in the early stages of development, provided it does not hinder usability~\citep{sast_for_review}. We first review research on vulnerable code propagation through tutorials, examples, and other learning resources. Next, we discuss usability and adoption challenges in current SAST practices. Finally, we examine repository-level automation through GitHub Actions and studies comparing SAST tools. We scope our work to text-based tutorials that either provide a code repository or lead to the creation of one, as opposed to video tutorials or generative AI output. Although generative code repositories are becoming common in development practices, we assume they require different methodological validation.

\subsection{Vulnerable Code Propagation and Learning Resources}

Developers learn from tutorials, examples, documentation, community resources, and existing code. Shared code snippets on Stack Overflow can change over time through community intervention \citep{jallow2025usenix}. Security resources can remain unchanged after publication \citep{selvaraj2024tosem}. Prior studies also found that insecure patterns in tutorials can persist and appear in downstream projects \citep{unruh2017leveraging,pathtraversal}. Research on code reuse shows that developers treat online resources as authoritative sources, which increases the effect of insecure examples \citep{copy-paste,meng2020copypaste}.

Recent studies also show that learners and developers learn from examples through AI~\citep{ardic2026genai}. However, research shows that the quality of generated code still needs validation. \citet{lekshmi2024examples} showed that LLM-generated explanations for code snippets are less readable than human-written explanations. In the context of whole repositories, \citet{jiang2026doc2feat} introduced Doc2Feat-Bench, in which software engineering agents receive documentation to create features, and reported failures in feature-development tasks involving existing code modules and documentation semantics. \citet{marceca2026agentic} evaluated Copilot, Cursor, and Windsurf on five complete web application tasks. The agents performed better on established application patterns but needed developers to provide guidance and iterative refinement during these tasks, with proper edge-case documentation. \citet{perry2023insecure} compared participants who developed code with and without an AI assistant. Participants who used the assistant produced less secure solutions and had greater confidence in the security of their solutions. \citet{copilot_security} evaluated Copilot and found exploitable vulnerabilities in generated code, including hardcoded credentials and injection flaws. \citet{pathtraversal} identified path-traversal vulnerabilities in 1,756 GitHub projects and examined how repeated instances of the pattern could be caused by LLM-generated code.

Django~\citep{django_security} and Spring~\citep{spring_securing_web} provide framework-level security guidelines and tutorials~\citep{django_cheatsheet}, but they provide security references separately rather than within beginner tutorials. Security tutorials also direct developers to broader resources such as OWASP. The OWASP Top 10 classifies major web application security risks such as injection, broken access control, and security misconfiguration~\citep{owasp_top_ten}.

\subsection{Security Practice Adoption and Alert Usability}

SASTs can identify security vulnerabilities in code during development, but their use alone might not lead developers to fix those vulnerabilities. Developers may need more details that match their task context, point to relevant code, explain the security concern, and propose a concrete fix.

Prior studies show where SAST alerts can be ineffective. \citet{sast_for_review} studied SAST alerts and found that they do not always align with the issue under review. Industry studies also show that benchmark performance does not fully capture practical SAST needs or how developers judge tool alerts \citep{comparison,li2025sast_eval}. \citet{shen2025codeql} applied CodeQL to open-source embedded software and showed that CodeQL could identify defects. However, developers still had to configure the analysis, validate alerts, and decide how to respond. \citet{suppressed_warning} studied SAST alerts suppressed by developers and showed that suppression reflected a developer's judgment about whether an alert deserved attention. Other work found that developers need relevant context, clear explanations, and concrete remediation guidance when they assess SAST alerts \citep{christakis2016developers,usability-criteria}.

Prior studies also examined the level of abstraction in SAST alerts. \citet{hermann2025securityfeatures} developed a taxonomy of implementation-level security features and examined their support across multiple security frameworks. They distinguished high-level security requirements from implementation mechanisms such as APIs, configuration options, and annotations. \citet{galanska2026recommendations} asked 598 IT professionals and cybersecurity students to evaluate alerts. Participants expressed a need for clarity, actionability, and confidence in the alerts.

AI-assisted security alert generation adds another context. \citet{sajadi2025security} and \citet{nath2026hammer} found concerns about the reliability and quality of LLM-generated security alerts, as well as the effort required to verify them. \citet{wong2026workforce} interviewed cybersecurity professionals and identified unreliable alerts and task-specific limitations as barriers to AI use, alongside concerns about data leakage. \citet{bappy2026reactive} studied professional software engineers and found that none of the participants in their coding sessions included security requirements in their initial prompts. This observation indicates a shift in security attention from code creation toward later review.

\subsection{GitHub Actions and SAST}

On the one hand, there is ample evidence of how developers create, discuss, maintain, and debug GitHub Actions workflows \citep{bouzenia2024resource,mastropaolo2024toward}. This evidence covers resource usage \citep{zhang2024developers}, causes of workflow failures \citep{zheng2026failures}, reruns \citep{huang2026reruns}, and automation reuse practices \citep{delicheh2026automation}.

Researchers have also characterized security weaknesses in GitHub CI workflows~\citep{characterizing_ci_workflows2022, ambush2024,continuous_intrusion2023,more_haste_less_speed2024,ci_plugins_ccs2024}. More specifically, they have examined GitHub Actions Marketplace risks~\citep{time_for_actions2025} and mitigation techniques \citep{revisiting_security_practices2025,mitigating_gha_security2024,ghast2022}. \citet{argus2023} built ARGUS to detect injection risks in workflows and actions through staged static taint analysis. Other studies examined permission misuse and permission hardening in GitHub Actions workflows~\citep{cosseter2026,granite2025}. Repository-level studies show similar differences in adoption and tool behavior. \citet{revisiting_security_practices2025} found that active repositories still used suboptimal GitHub Actions security practices. Moreover, there was substantial variation in adoption, with implementation rates ranging from 0.6\% to 52.9\% across five practices~\citep{action_required2026}. Researchers linked this variation to awareness, applicability, and setup barriers. Available GitHub Security Actions vary in scope, detection, and reported results~\citep{unpacking_scanners2026}. These results show that developer awareness and decisions about tool selection and configuration affect the security information developers receive.

On the other hand, various studies have investigated SASTs and the characteristics of their alerts. \citet{nocera2026python} analyzed Python projects and commits with SonarQube to study the type, severity, and persistence of security issues. \citet{zhang2025julia} examined Julia programs from open-source projects and developed a static-analysis approach to characterize security weaknesses in the Julia ecosystem. Other studies compared alerts across analysis tools. \citet{diangelo2024evolution} compared automated weakness-detection tools on Ethereum bytecode and found differences in detected weaknesses and agreement between tools. \citet{comparison,li2025sast_eval} studied how practitioners evaluate SAST tools and found that practitioners use a combination of tools for better performance. \citet{unpacking_scanners2026} compared GitHub Actions workflow security scanners and found differences in scanner scope, detection, and reported results.

These studies examine the security of GitHub Actions workflows and differences in the findings reported by SAST tools. Our study focuses on the alerts generated by default GitHub Security Actions configurations in Django and Spring tutorial repositories, examining their relevance to tutorial code and the remediation guidance they provide.

We do not assume that developers use step-by-step tutorials as their only way to learn how to code with a given technology. We treat tutorial repositories as one class of mature learning artifacts. Our study focuses on how security alerts can be helpful in the early stages of development when learning takes place. Prior work examines the need for security due to vulnerability propagation and a lack of security knowledge, as well as industry practices such as workflow security, SAST adoption, and usability. We examine how current industry practices such as the use of GitHub Security Actions present security alerts for Django and Spring tutorial repositories. We characterize the alerts in the context of tutorial repositories.

\begin{table*}[t]
\centering
\caption{Properties, Rationale, and Statistical Analysis.}
\label{tab:metric-definitions}
\small
\setlength{\tabcolsep}{4pt}
\renewcommand{\arraystretch}{1.15}
\resizebox{\textwidth}{!}{%
\begin{tabular}{p{2.2cm} p{4.0cm} p{4.0cm} p{4.0cm}}
\toprule
\textbf{Property} &
\textbf{Rationale} &
\textbf{Operational Definition} &
\textbf{Statistical Approach} \\
\midrule
\textbf{Alert Volume} &
High alert volume increases cognitive load and can lead learners to ignore alerts entirely. &
\( |A| \): Total count of alerts reported for a single tutorial. &
\textbf{Wilcoxon Signed-Rank:} To test whether different tools produce different alert volumes. \\
\midrule
\textbf{Alert Scope} &
Distinguishes between alerts targeting code the learner wrote versus library or generated code. &
Categorical classification:\newline
\( A_{\mathrm{scope}} \in
\{\text{Tutorial},\text{Non-Tutorial}\} \) &
\textbf{Descriptive:} Ratio of alerts in scope. \\
\midrule
\textbf{Rule Diversity} &
High diversity forces learners to context-switch between many different security concepts. &
\( |R| \): Count of distinct rules triggered per tutorial. &
\textbf{Descriptive:} Differences between tool configurations over all tutorials. \\
\midrule
\textbf{Rule Concentration} &
High concentration indicates that alerts focus on a small set of recurring rules rather than a wider set of rules. &
\(\displaystyle
\frac{\sum_{r \in \text{Top-5 }R}\operatorname{count}(A_r)}
{|A|}
\) &
\textbf{Descriptive:} To examine whether alerts concentrate on a small set of recurring rules. \\
\midrule
\textbf{CWE Coverage} &
Describes the security weaknesses associated with the reported alerts. &
\( |\operatorname{CWE}(A)| \): Number of distinct CWE identifiers; number of alerts associated with each CWE. &
\textbf{Descriptive:} CWE frequencies and alerts without a CWE identifier. \\
\midrule
\textbf{OWASP Coverage} &
Ensures that reducing alert volume does not result in overlooking categories from the OWASP Top 10:2025. &
CWE-to-OWASP mapping:\newline
\( a \rightarrow \operatorname{CWE}(a)
\rightarrow \mathrm{OWASP}_{2025},\quad a \in A \) &
\textbf{Descriptive:} Coverage of OWASP 2025 categories by tool configurations. \\
\midrule
\textbf{Alert Length} &
Lengthy descriptions increase reading time and cognitive load and can obscure remediation. &
\(\displaystyle
\mu_{\mathrm{words}}
=
\frac{1}{|A|}
\sum_{a \in A}
\operatorname{word\_count}(a)
\) &
\textbf{Descriptive:} Differences between tool configurations over all tutorials. \\
\midrule
\textbf{Concrete-Reference Rate} &
Indicates whether selected exported text names a framework or code construct that the reader can locate in the application. &
\(\displaystyle
CR(A)=\frac{1}{|A|}\sum_{a\in A}I_C(m_a)
\), where \(I_C\) indicates a construct-lexicon match. &
\textbf{Descriptive:} Percentage of selected messages that match the construct lexicon. \\
\midrule

\textbf{Remediation-Specific Rate} &
Indicates whether selected exported text contains
both a concrete construct and an action term. &
\(\begin{aligned}
RS(A) &= \\\frac{1}{|A|} 
&\quad \sum_{a\in A} I_C(m_a)I_V(m_a)
\end{aligned}\)

where \(I_V\) indicates an action-verb match. &
\textbf{Descriptive:} Percentage of selected
messages that satisfy both lexical conditions. \\

\bottomrule
\end{tabular}%
}
\end{table*}

\section{Methodology}
\label{sec:methodology}

We conducted a comparative study of the security alerts that CodeQL, Semgrep, and SonarCloud produced for tutorial-based web application repositories. \emph{GitHub Security Actions} report alerts through GitHub code scanning or the tool dashboard. We examined the distribution, location, rule metadata, and message content of the alerts. We analyzed properties of the security alerts that would be relevant to tutorial repositories. Section~\ref{ss:metric-definitions} summarizes the properties we studied, which are also summarized in Table~\ref{tab:metric-definitions}.

\subsection{Tutorial Corpus Construction}
\label{ss:tutorial-corpus}

We collected 60 web application tutorial repositories targeting Django (37) and Spring (23), two mature frameworks with substantial use in professional development. According to the 2025 Stack Overflow Developer Survey~\citep{stackoverflow20252}, Spring accounts for 15.6\% of professional web framework usage and Django accounts for 11.7\%, for a combined 27.3\% among respondents. We selected these frameworks because they represent established ecosystems with extensive learning material and because they support realistic server-side web application development.

We used DuckDuckGo and GitHub searches in July 2026 to identify tutorials. For DuckDuckGo, we considered the first 50 results in worldwide search. Our queries included \emph{``best Django tutorial''}, \emph{``Django complete guide''}, \emph{``Spring Boot tutorial''}, and \emph{``Spring Boot complete guide''}. For GitHub, we searched repository names, descriptions, and README files with the terms \emph{tutorial}, \emph{guide}, \emph{course}, \emph{example}, \emph{starter}, and \emph{sample application}, in that order \footnote{GitHub query form : \code{Django \textless term\textgreater{} in:name,description,readme language:Python stars:\textgreater=500} and \code{Spring \textless term\textgreater{} in:name,description,readme language:Java stars:\textgreater=500}}. 

We included tutorials that produced a complete, runnable web application. A tutorial met this criterion if it either linked to a public source code repository or provided enough step-by-step instructions for us to reconstruct the application. For GitHub learning repositories, we also required more than 500 stars and at least one commit on or after January 1, 2024. We excluded repositories that contained only isolated snippets, framework configuration examples, libraries, templates without a runnable application, or non-English learning material. When a repository existed, we forked it to preserve a stable snapshot. Otherwise, we manually reconstructed the application from the tutorial instructions and archived it in a dedicated repository. We report the references to all studied repositories at the end of this paper, and the data collection process in our replication package~\citep{psp_replication_package}.

\subsection{GitHub Security Actions Configuration and Execution}
\label{ss:industrial-sasts}

We ran three tools using their default configuration on GitHub Actions-based workflows: CodeQL, SonarCloud with Security Hotspots, and Semgrep. We selected these tools based on their use in prior empirical studies and their ability to run at the repository level~\citep{comparison,pathtraversal}.

The tools represent distinct static analysis approaches. CodeQL uses a database-driven model that converts source code into a relational representation for whole-program analysis. It maps security properties as declarative queries and tracks data flows from sources to sinks through global data-flow and control-flow relations~\citep{codeql-dataflow}. SonarCloud uses a hybrid analysis engine that combines syntactic checks with semantic reasoning. For Java projects, its analyzers use symbolic execution to explore feasible execution paths and identify path-dependent defects~\citep{sonar-symbolic-execution}. Semgrep primarily uses syntax-tree pattern rules that define structural code patterns and map rule definitions directly to code matches~\citep{semgrep-patterns}. Together, these tools cover different static analysis techniques and provide complementary views of security alerts in tutorial repositories.

We configured each tool with a workflow file under \texttt{.github/workflows/}. We used default workflow templates when possible because they reflect a low effort configuration path for security beginners. We avoided project-specific tuning because security beginners may not know which rules to select, which paths to exclude, or how to adjust SAST thresholds.

We ran CodeQL with the default \texttt{github/codeql-action/init} and ~\texttt{github /codeql-action /analyze} steps and did not specify a custom query suite. CodeQL therefore used its default security query suite for each supported language. We ran Semgrep with \texttt{semgrep ci} and did not provide custom \texttt{--config} flags. In this mode, Semgrep used the policy and rule configuration associated with the Semgrep project; we did not add tutorial-specific rules, suppressions, or custom rule files. We ran SonarCloud through the official SonarSource GitHub Actions workflow for SonarCloud and invoked the SAST through Maven or Gradle according to each tutorial build system. SonarCloud used its default \emph{Sonar way} quality profile.

\subsection{Alert Extraction and Data Collection}
\label{ss:alert-extraction}

We extracted alerts programmatically from the outputs produced by the GitHub Actions workflow runs. We collected CodeQL and Semgrep alerts through the GitHub code scanning APIs~\citep{gitapi}. We retrieved SonarCloud and Security Hotspots alerts through the SonarCloud Web API~\citep{sonarapi}. For each tutorial repository, we retrieved all alerts and recorded the tool name, rule identifier, file path, line range, severity when available, and short description. We treated SARIF\footnote{\href{https://sarifweb.azurewebsites.net/}{Static Analysis Results Interchange Format (SARIF)}} artifacts produced during CI execution as the canonical source for CodeQL and Semgrep. We exported alerts in the SARIF to support uniform downstream processing. 

We treated each exported alert record as an observation for properties that depend on individual alerts, such as alert scope, alert length, CWE coverage, and OWASP category coverage. Multiple records could share the same tool, rule identifier, file path, and line range, and we retained these records in the analysis. For properties associated with an entire tutorial, such as alert volume and rule diversity, we aggregated the records at the tutorial level. We retained alerts reported by different tools on the same line of code as separate observations.

\subsection{Alert Properties}
\label{ss:rq1-measures}
\label{ss:metric-definitions}
Table~\ref{tab:metric-definitions} shows the nine properties we measured. 

\textbf{Alert volume} counts how many alerts a learner receives during a tutorial. We counted every reported alert because a learner sees the full set of alerts before they can judge relevance or correctness. We compared paired tool configurations with Wilcoxon signed-rank tests. We also reported Cliff's delta as a descriptive effect size, comparing the distributions of per-repository alert counts between configurations. We discuss false positives in Section~\ref{sec:limitations}.

\textbf{Alert scope} locates each alert in the repository. We classified every alert as tutorial or non-tutorial based on its file path. Tutorial scope covers code that relates to the tutorial task. Non-tutorial scope covers four kinds of files: (i) dependency-installed code, such as \texttt{.venv}, \texttt{site-packages}, and \texttt{node\_modules}\footnote{Some Spring and Django tutorials include \codesm{node\_modules} because they contain frontend asset dependencies, build tooling, or client-side resources. We treated these files as dependency-installed code.}; (ii) generated framework artifacts, such as Django \texttt{migrations}; (iii) build outputs and caches, such as \texttt{\_\_pycache\_\_}, compiled files, and generated target directories; and (iv) repository files unrelated to the tutorial programming task, such as CI workflow files, dependency manifests (e.g., \texttt{package-lock.json}, \texttt{requirements.txt}, and \texttt{pom.xml}), and build metadata (e.g., \texttt{build.gradle}). We kept test files in tutorial scope when the repository included them. We applied the classification through deterministic path-matching rules. We then measured alert scope as the ratio of tutorial-scoped alerts to all reported alerts. For Table~\ref{tab:tutorial-repositories}, we also separately counted alerts in GitHub Actions workflow files and dependency manifests or build metadata as subsets of non-tutorial alerts.

\textbf{Rule diversity} and \textbf{rule concentration} describe the spread of the rules among the alerts. Rule diversity counts how many distinct rules a tutorial triggers. Rule concentration measures whether the alerts cluster around a few recurring rules (we consider the five most frequent). Together, the two measures show whether a learner encounters a focused set of recurring concepts or a wide set of unrelated alerts. We compared both measures across tool configurations and across all tutorials.

\textbf{CWE coverage} describes the Common Weakness Enumeration (CWE) identifiers associated with the reported alerts. We extracted CWE identifiers from the exported alert records and normalized their format. We counted the number of alerts with and without CWE identifiers, the number of distinct CWEs, and the number of alerts associated with each CWE. When an alert contained multiple CWE identifiers, we counted it once for each distinct CWE. We compared the CWE frequencies across tool configurations.

\textbf{OWASP coverage} describes which security categories appear in the triggered rules. We mapped the recorded CWE identifiers to OWASP Top 10:2025 categories using the CWE lists published for the ten categories. An alert was counted once in each corresponding category when its CWEs mapped to more than one category. We recorded alerts without a corresponding OWASP 2025 category separately. We compared category coverage across tool configurations. This property shows whether a lower alert volume also narrows the range of security concepts in tutorial alerts.

We examined alert messages with three properties: \textbf{Alert length}, the \textbf{concrete-reference rate}, and the \textbf{remediation-specific rate}. We drew all three from the alert message text, which we took as the first non-empty field among the alert title, the rule description, and the rule name. Alert length measures the mean word count of that text. The concrete-reference rate measures the ratio of messages that name a framework or code construct. We matched each message against a lexicon of named constructs that spans both frameworks' security properties: decorators and annotations such as \texttt{login\_required} and \texttt{@PreAuthorize}; framework settings and configuration keys such as \texttt{DEBUG} and \texttt{SECRET\_KEY}; and API or method calls such as \texttt{HttpResponse} and explicit call forms.

We measured the remediation-specific rate as the ratio of messages that state a specific local fix. We jointly applied two conditions. First, a message had to contain an action verb from a fixed list, such as \texttt{set}, \texttt{disable}, \texttt{configure}, \texttt{replace}, \texttt{sanitize}, or \texttt{escape}. Second, the message had to contain a concrete construct from the lexicon above, so the action had an object that learners could verify. We labeled a message as remediation-specific only when it met both conditions. For example, ``set \texttt{DEBUG} to false,'' is remediation-specific whereas a message such as ``make sure this endpoint is safe'' is not.

We applied all measures through deterministic pattern matching, so the classification can be repeated identically across tutorials and tool configurations. We list the full path rules, CWE identifiers, CWE-to-OWASP 2025 mappings, construct lexicon, and action-verb lexicon in the replication package~\citep{psp_replication_package}.

\subsection{Limitations}
\label{sec:limitations}

We analyzed 60 Django and Spring tutorial repositories (Section~\ref{ss:tutorial-corpus}). The corpus captures only part of the tutorial ecosystem, which differs across platforms, languages, and instructional goals. We therefore make no claim of statistical representativeness.

We decided to use the defaults that the tools supplied through GitHub Actions Marketplace.\footnote{GitHub Actions Marketplace: \url{https://github.com/marketplace?type=actions}} We did not tune the configurations for tutorials to represent the alerts that a security beginner could obtain through the workflow templates and profiles. The defaults depend on the tool and its configuration. For example, \texttt{semgrep ci} used the Semgrep project's policy, and policies can vary between installations. Our alerts apply to the workflows and policies we recorded. Changes to repositories, rules, analysis engines, or services can produce different output. The replication package provides the workflows, versions, scan dates, rule snapshots, and exports that we used~\citep{psp_replication_package}. Expert developers can adjust rules and severity thresholds, but security beginners may not know which changes to make. Our results therefore characterize default alerts in tutorial contexts rather than optimized tool performance. Our rule-level analysis includes only alerts and rules that appeared during scanning. We did not examine untriggered rules from each tool. The mapping therefore reflects the security-relevant code patterns present in our tutorial corpus, not each tool's full possible coverage.

Our tutorial-scoping relies on deterministic path rules. This deterministic process supports repeatability, but different definitions of tutorial code could change the proportion of alerts classified as tutorial-scoped for some repositories. We also use deterministic lexicons as proxies for actionability. These metrics show whether an alert names a local construct or repair action, but they do not measure how much background knowledge a learner needs. Future work could test whether these alert properties affect how learners interpret alerts and modify code.

Static analysis tools can produce false positives~\citep{esposito2024extensive}. We retained all alerts from the configurations because developers see the full GitHub Actions output before they judge scope, relevance, or actionability. We designed the manual review to check the correctness of alerts associated with the most-triggered rules, rather than to estimate each tool's overall precision. Rare rules may show different false-positive behavior. We manually reviewed 80 alerts, with 20 alerts per tool. For each alert, we examined the rule, affected code, and references. Two reviewers coded each alert as a true positive or false positive. We classified 76 alerts as true positives and 4 as false positives (Table~\ref{tab:false-positive-distribution}). The review produced one disagreement, with Cohen's $\kappa = 0.95$. It concerned a Django REST Framework throttling alert: one reviewer viewed authentication as sufficient mitigation, while the other viewed missing throttling as valid.

\begin{table}[t]
\centering
\caption{Review of alerts from the four configurations. TP and FP denote true and false positives.}
\label{tab:false-positive-distribution}
\footnotesize
\begin{tabular}{lrrr}
\toprule
\textbf{Configuration} & \textbf{TP} & \textbf{FP} & \textbf{Total} \\
\midrule
CodeQL & 20 & 0 & 20 \\
Semgrep & 17 & 3 & 20 \\
SonarCloud & 19 & 1 & 20 \\
SonarCloud Security Hotspots & 20 & 0 & 20 \\
\midrule
\textbf{Total} & \textbf{76} & \textbf{4} & \textbf{80} \\
\bottomrule
\end{tabular}
\end{table}

Our message analysis depends on the metadata fields we selected. We used the alert title, rule description, and rule name. We calculated the rates over alerts, so repeated messages affected the results. These measures describe the occurrence of words and constructs in the SARIF output we collected; they do not measure the knowledge a reader needs to interpret an alert or assess the usability of the tool interface.

\begin{figure*}[!htbp]

\centering

\includegraphics[width=.87\linewidth]{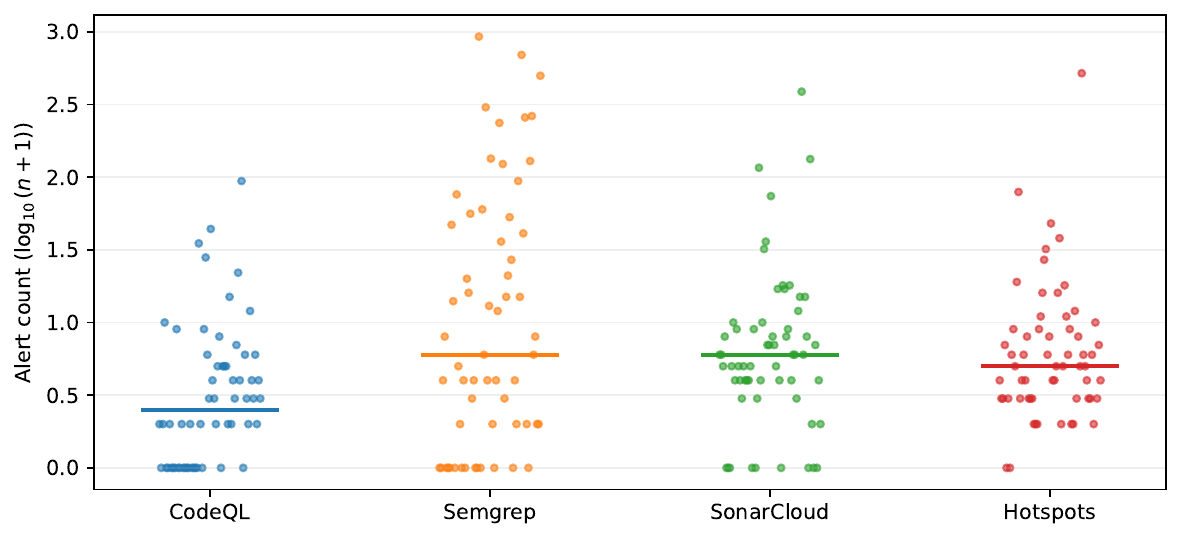}

\caption{Alert counts per repository for the four default SASTs. Each point represents a repository; horizontal marks indicate the medians.}

\label{fig:alert-volume-distr}

\end{figure*}
\section{Results}

\label{sec:results}

Our overarching research question is \emph{To what extent are alerts from default GitHub Security Actions configurations applicable to tutorials?} To answer this question, we ran four GitHub Security Actions configurations (CodeQL, Semgrep, SonarCloud, and Hotspots) on 60 tutorial repositories and analyzed 6,609 alerts. We treated each repository $\times$ tool combination as one unit of analysis because a learner receives the output of one tool at a time. We report the results across properties defined in Section~\ref{ss:metric-definitions} and summarized in Table~\ref{tab:metric-definitions}.

\subsection{Dataset Overview}
\label{ss:tutorial-corpus-overview}

Our dataset contains 37 Django and 23 Spring repositories. Repository sizes varied, Table~\ref{tab:tutorial-repositories} provides the lines of code (LOC) measurements\footnote{We measured LOC using \texttt{cloc} (\url{https://github.com/AlDanial/cloc}). The tool counts non-blank, non-comment lines in recognized file types, including documentation formats such as Markdown and text files.} and security alert counts for all 60 repositories. The Django repositories LOC ranged between 184 and 87,803, with a mean of 6,548.6 and a median of 1,874 LOC. For Spring repositories it is between 151 and 1,140,914, with a mean of 63,952.5 and a median of 1,322 LOC. We observed that a small number of large repositories influenced the averages, particularly within the Spring repositories. 

\subsection{Alert Volume}

\label{ss:AlertVolume}

\emph{Alert volume} captures how many alerts a learner encounters when running GitHub Security Actions on a tutorial repository. Figure~\ref{fig:alert-volume-distr} summarizes the distribution of alert counts across the 60 repositories for each configuration. Table~\ref{tab:tutorial-repositories} reports the total number of alerts, the number of repositories with alerts, and the mean, median, and maximum alert count for each configuration. Semgrep reported the largest number of alerts (4,206), while Hotspots reported alerts in the largest number of repositories (58 of 60). For all four configurations, the mean alert count exceeded the median, with the largest difference for Semgrep (70.1 compared with 5). Some repository $\times$ configuration combinations had few or no reported alerts (see also Table~\ref{tab:tutorial-repositories}). We compared alert counts between configurations using paired Wilcoxon signed-rank tests, as described in Section~\ref{ss:metric-definitions}. Semgrep reported higher alert counts than CodeQL, with a small effect size (Cliff's $\delta = 0.32$). Hotspots also reported higher alert counts than CodeQL, with a moderate effect size (Cliff's $\delta = 0.39$).

The maximum Semgrep count (925 alerts) occurred in \texttt{GH-SB-FullStack}. In comparison, the maximum counts for CodeQL (93), SonarCloud (386), and Hotspots (517) all occurred in \texttt{GH-SB-JournalDev}. Across the four configurations, \texttt{GH-SB-JournalDev} had 25.5\% of the alerts, followed by \texttt{GH-SB-FullStack} with 16.4\%. The five repositories with the largest combined alert counts was 60.0\% of all alerts. For only one repository, \texttt{SB-Geeks}, none of the four configurations reported an alert.

We examined repository size to characterize this variation. The LOCs in Table~\ref{tab:tutorial-repositories} show that most repositories had relatively small codebases, while a few were substantially larger. The largest repository, \texttt{GH-SB-FullStack}, contains a collection of documented Spring Boot examples covering configuration, REST APIs, security, testing, and web applications. However, \texttt{GH-SB-JournalDev} had more alerts than \texttt{GH-SB-FullStack}, although it had fewer LOC. Similarly, \texttt{GH-SB-Data} contained a relatively large codebase but had few reported alerts. Thus, more LOCs do not necessarily mean more alerts. The median combined alert count per repository was higher for Django than for Spring. The Spring total was concentrated in a small number of repositories, particularly \texttt{GH-SB-JournalDev} and \texttt{GH-SB-FullStack}.

We also examined the files in which alerts were reported. In Table~\ref{tab:tutorial-repositories}, 11 Spring repositories had alerts in dependency files and build metadata, compared with 27 Django repositories. These 11 Spring repositories had 1,595 dependency-file and build-metadata alerts, more than twice the 736 reported for the 27 Django repositories. For example, \texttt{GH-SB-SecurityAngular} had 471 dependency-file alerts. This repository contains several Angular application examples, including separate \texttt{package-lock.json} files recording JavaScript package dependencies. In comparison, the Django repository \texttt{GH-DJ-React} had 117 dependency-file and build-metadata alerts and includes a \texttt{pnpm-lock.yaml} file for its React frontend.

Alert counts also varied between repositories and concentrated in a small number of cases. The examples involving frontend applications and dependency manifests show the different kinds of repository files in which alerts occurred. 
Thus, alert volume is not an indication of whether the tutorial code is secure or insecure.

\begingroup

\scriptsize

\setlength{\tabcolsep}{2.0pt}

\renewcommand{\arraystretch}{1.10}

\begin{longtable}{@{}p{2.5cm} r r r r r r r r r r r@{}}

\caption{Tutorial repositories grouped by framework, with per-framework and overall alert volume and alert scope. Work. counts alerts in GitHub Actions workflow files across all four configurations; Dep. counts dependency manifests and build metadata; CWE counts alerts with at least one reported or rule-derived Common Weakness Enumeration identifier across all four configurations.}

\label{tab:tutorial-repositories}\\

\toprule

\multicolumn{2}{c}{} & \multicolumn{5}{c}{\textbf{Alert volume}} & \multicolumn{5}{c}{\textbf{Alert scope and CWE}} \\

\cmidrule(lr){3-7}\cmidrule(lr){8-12}

Repository & LOC & CQL & Sem. & Sonar & Hot. & All & Tut. & Other & Work. & Dep. & CWE \\

\midrule

\endfirsthead



\multicolumn{2}{c}{} & \multicolumn{5}{c}{\textbf{Alert volume}} & \multicolumn{5}{c}{\textbf{Alert scope and CWE}} \\

\cmidrule(lr){3-7}\cmidrule(lr){8-12}

Repository & LOC & CQL & Sem. & Sonar & Hot. & All & Tut. & Other & Work. & Dep. & CWE \\

\midrule

\endhead

\midrule

\multicolumn{12}{r}{\emph{Continued on next page}}\\

\endfoot

\bottomrule

\endlastfoot

\addlinespace[3pt]

\multicolumn{12}{@{}l}{\textbf{Django (37 repositories)}}\\

\midrule

\href{https://github.com/research-projects-all/realpython-main-tutorial-8}{DJ-RealPython1} & 359 & 0 & 46 & 4 & 5 & 55 & 7 & 48 & 3 & 45 & 49 \\

\href{https://github.com/research-projects-all/realpython-main-tutorial-9}{DJ-RealPython2} & 1,874 & 0 & 13 & 9 & 8 & 30 & 14 & 16 & 3 & 13 & 28 \\

\href{https://github.com/research-projects-all/django-girls-main-tutorial-10}{DJ-Girls} & 416 & 0 & 0 & 3 & 4 & 7 & 4 & 3 & 3 & 0 & 2 \\

\href{https://github.com/research-projects-all/tango_with_django_19-main-tutorial-11}{DJ-Tango} & 1,885 & 8 & 75 & 8 & 18 & 109 & 54 & 55 & 3 & 52 & 97 \\

\href{https://github.com/research-projects-all/learndjango-rest-framework-tutorial-main-tutorial-13}{DJ-LearnDRF} & 1,014 & 0 & 4 & 4 & 78 & 86 & 3 & 83 & 3 & 3 & 8 \\

\href{https://github.com/research-projects-all/django-restframework-main-tutorial-14}{DJ-DRFQuick} & 262 & 0 & 1 & 3 & 2 & 6 & 3 & 3 & 3 & 0 & 4 \\

\href{https://github.com/research-projects-all/django-tutorial-dev-main-tutorial-15}{DJ-TutorialDev} & 230 & 1 & 0 & 2 & 3 & 6 & 2 & 4 & 4 & 0 & 3 \\

\href{https://github.com/research-projects-all/geeksforgeeks-django-main-tutorial-16}{DJ-TodoGeeks} & 347 & 0 & 3 & 4 & 5 & 12 & 9 & 3 & 3 & 0 & 6 \\

\href{https://github.com/research-projects-all/django-academy-main-tutorial-17}{DJ-Academy} & 291 & 0 & 0 & 3 & 3 & 6 & 3 & 3 & 3 & 0 & 3 \\

\href{https://github.com/research-projects-all/sibtc-django-beginners-guide-main-tutorial-18}{DJ-SIBTC} & 1,924 & 0 & 19 & 3 & 7 & 29 & 15 & 14 & 3 & 11 & 22 \\

\href{https://github.com/research-projects-all/pythontutorial-main-tutorial-19}{DJ-PythonTutorial} & 235 & 0 & 15 & 3 & 2 & 20 & 2 & 18 & 3 & 15 & 18 \\

\href{https://github.com/research-projects-all/djangobin-main-tutorial-20}{DJ-OverIQ} & 2,460 & 1 & 55 & 4 & 2 & 62 & 20 & 42 & 3 & 39 & 59 \\

DJ-Geeks & 213 & 0 & 0 & 0 & 1 & 1 & 0 & 1 & 1 & 0 & 1 \\

\href{https://github.com/research-projects-all/Try-Django}{GH-DJ-TryDjango} & 816 & 1 & 0 & 3 & 10 & 14 & 10 & 4 & 4 & 0 & 5 \\

\href{https://github.com/research-projects-all/channels-examples}{GH-DJ-Channels} & 1,144 & 0 & 59 & 9 & 15 & 83 & 27 & 56 & 3 & 53 & 76 \\

\href{https://github.com/research-projects-all/cookiecutter-django}{GH-DJ-Cookiecutter} & 2,593 & 8 & 5 & 31 & 26 & 70 & 29 & 41 & 39 & 2 & 47 \\

\href{https://github.com/research-projects-all/django-allauth}{GH-DJ-Allauth} & 41,247 & 27 & 301 & 35 & 31 & 394 & 240 & 154 & 14 & 140 & 372 \\

\href{https://github.com/research-projects-all/django-analytical}{GH-DJ-Analytical} & 2,465 & 5 & 3 & 6 & 5 & 19 & 6 & 13 & 12 & 1 & 13 \\

\href{https://github.com/research-projects-all/django-celery}{GH-DJ-Celery} & 4,163 & 2 & 12 & 6 & 7 & 27 & 13 & 14 & 7 & 7 & 23 \\

\href{https://github.com/research-projects-all/django-cms}{GH-DJ-CMS} & 87,803 & 43 & 133 & 73 & 47 & 296 & 175 & 121 & 89 & 32 & 203 \\

\href{https://github.com/research-projects-all/django-configurations}{GH-DJ-Config} & 1,126 & 3 & 1 & 7 & 3 & 14 & 2 & 12 & 12 & 0 & 8 \\

\href{https://github.com/research-projects-all/django-controlcenter}{GH-DJ-ControlCenter} & 2,323 & 2 & 0 & 6 & 3 & 11 & 1 & 10 & 9 & 1 & 5 \\

\href{https://github.com/research-projects-all/django-locallibrary-tutorial}{GH-DJ-LocalLibrary} & 2,112 & 1 & 3 & 4 & 4 & 12 & 6 & 6 & 5 & 1 & 9 \\

\href{https://github.com/research-projects-all/django-organizations}{GH-DJ-Organizations} & 4,009 & 4 & 11 & 16 & 15 & 46 & 30 & 16 & 15 & 1 & 31 \\

\href{https://github.com/research-projects-all/django-oscar}{GH-DJ-Oscar} & 43,138 & 7 & 235 & 3 & 37 & 282 & 221 & 61 & 1 & 60 & 279 \\

\href{https://github.com/research-projects-all/django-project-template}{GH-DJ-Template} & 184 & 0 & 35 & 0 & 1 & 36 & 1 & 35 & 0 & 35 & 36 \\

\href{https://github.com/research-projects-all/django-react-boilerplate}{GH-DJ-React} & 10,345 & 4 & 122 & 17 & 4 & 147 & 16 & 131 & 14 & 117 & 131 \\

\href{https://github.com/research-projects-all/django-rest-framework-gis}{GH-DJ-DRF-GIS} & 1,392 & 4 & 2 & 16 & 17 & 39 & 8 & 31 & 30 & 1 & 24 \\

\href{https://github.com/research-projects-all/django-rosetta}{GH-DJ-Rosetta} & 3,035 & 4 & 14 & 7 & 10 & 35 & 25 & 10 & 9 & 1 & 28 \\

\href{https://github.com/research-projects-all/django-summernote}{GH-DJ-Summernote} & 3,285 & 1 & 20 & 8 & 5 & 34 & 30 & 4 & 3 & 1 & 27 \\

\href{https://github.com/research-projects-all/django-two-factor-auth}{GH-DJ-2FA} & 3,719 & 14 & 52 & 17 & 8 & 91 & 75 & 16 & 15 & 1 & 76 \\

\href{https://github.com/research-projects-all/django-twoscoops-project}{GH-DJ-TwoScoops} & 435 & 1 & 26 & 3 & 1 & 31 & 6 & 25 & 3 & 22 & 28 \\

\href{https://github.com/research-projects-all/django-upgrade}{GH-DJ-Upgrade} & 4,702 & 3 & 0 & 5 & 1 & 9 & 0 & 9 & 9 & 0 & 4 \\

\href{https://github.com/research-projects-all/docker-django-example}{GH-DJ-Docker} & 738 & 2 & 3 & 5 & 11 & 21 & 13 & 8 & 5 & 3 & 17 \\

\href{https://github.com/research-projects-all/eCommerce}{GH-DJ-ECommerce} & 8,819 & 21 & 93 & 11 & 7 & 132 & 70 & 62 & 3 & 43 & 122 \\

\href{https://github.com/research-projects-all/lithium}{GH-DJ-Lithium} & 734 & 0 & 40 & 5 & 5 & 50 & 11 & 39 & 3 & 36 & 46 \\

\href{https://github.com/research-projects-all/rest-framework-tutorial}{GH-DJ-DRF} & 462 & 2 & 1 & 7 & 3 & 13 & 4 & 9 & 9 & 0 & 8 \\

\midrule

\multicolumn{1}{@{}l}{\quad\emph{Total}} & 242,299 & 169 & 1,402 & 350 & 414 & 2,335 & 1,155 & 1,180 & 351 & 736 & 1,918 \\

\multicolumn{1}{@{}l}{\quad\emph{Mean}} & 6,548.6 & 4.6 & 37.9 & 9.5 & 11.2 & 63.1 & 31.2 & 31.9 & 9.5 & 19.9 & 51.8 \\

\multicolumn{1}{@{}l}{\quad\emph{Median}} & 1,874.0 & 1.0 & 12.0 & 5.0 & 5.0 & 31.0 & 10.0 & 16.0 & 3.0 & 2.0 & 24.0 \\

\multicolumn{1}{@{}l}{\quad\emph{SD}} & 16,672.4 & 8.7 & 65.8 & 13.1 & 15.4 & 87.5 & 57.8 & 37.6 & 15.6 & 32.4 & 79.8 \\

\multicolumn{1}{@{}l}{\quad\emph{Max}} & 87,803 & 43 & 301 & 73 & 78 & 394 & 240 & 154 & 89 & 140 & 372 \\

\multicolumn{1}{@{}l}{\quad\emph{Repos. $>0$}} & 37 & 24 & 30 & 35 & 37 & 37 & 35 & 37 & 36 & 27 & 37 \\

\midrule\midrule

\addlinespace[3pt]

\multicolumn{12}{@{}l}{\textbf{Spring (23 repositories)}}\\

\midrule

\href{https://github.com/research-projects-all/gs-spring-boot--main-tutorial-1-}{SB-GettingStarted} & 291 & 1 & 0 & 5 & 3 & 9 & 0 & 9 & 5 & 4 & 4 \\

\href{https://github.com/research-projects-all/gs-rest-service-main-tutorial-2}{SB-REST} & 307 & 0 & 0 & 5 & 2 & 7 & 0 & 7 & 3 & 4 & 2 \\

\href{https://github.com/research-projects-all/beuldang-main-tutorial-3}{SB-BaeldungBoot} & 532 & 1 & 3 & 4 & 2 & 10 & 10 & 0 & 0 & 0 & 8 \\

\href{https://github.com/research-projects-all/bealdung-main-tutorial-4}{SB-BaeldungMVC} & 1,342 & 9 & 7 & 7 & 6 & 29 & 29 & 0 & 0 & 0 & 23 \\

\href{https://github.com/research-projects-all/geekforgeeks-main-tutorial-5}{SB-Geeks}\textsuperscript{$\dagger$} & 213 & 0 & 0 & 0 & 0 & 0 & 0 & 0 & 0 & 0 & 0 \\

\href{https://github.com/research-projects-all/doc-springio-main-tutorial-6}{SB-DocsFirst} & 151 & 0 & 0 & 0 & 2 & 2 & 0 & 2 & 2 & 0 & 2 \\

\href{https://github.com/research-projects-all/gs-serving-web-content-main-tutorial-7}{SB-ServingWeb} & 171 & 1 & 0 & 0 & 0 & 1 & 0 & 1 & 1 & 0 & 1 \\

\href{https://github.com/research-projects-all/SpringBootHelloWorldExample-main-tutorial-21}{SB-Java2Blog} & 243 & 0 & 2 & 0 & 2 & 4 & 2 & 2 & 2 & 0 & 4 \\

\href{https://github.com/research-projects-all/springboot-tutorial-firstjava-main-tutorial-23}{SB-FirstJava} & 408 & 0 & 3 & 8 & 1 & 12 & 12 & 0 & 0 & 0 & 11 \\

\href{https://github.com/research-projects-all/E-commerce-project-springBoot}{GH-SB-ECommerce} & 1,322 & 0 & 0 & 2 & 1 & 3 & 2 & 1 & 1 & 0 & 3 \\

\href{https://github.com/research-projects-all/SpringBoot-Projects-FullStack}{GH-SB-FullStack} & 1,140,914 & 34 & 925 & 115 & 8 & 1,082 & 468 & 614 & 2 & 612 & 1038 \\

\href{https://github.com/research-projects-all/dubbo-spring-boot-project}{GH-SB-Dubbo} & 4,459 & 6 & 1 & 2 & 2 & 11 & 6 & 5 & 4 & 1 & 9 \\

\href{https://github.com/research-projects-all/initializr}{GH-SB-Initializr} & 43,809 & 3 & 14 & 14 & 4 & 35 & 16 & 19 & 6 & 0 & 25 \\

\href{https://github.com/research-projects-all/journaldev}{GH-SB-JournalDev} & 122,574 & 93 & 692 & 386 & 517 & 1,688 & 1,657 & 31 & 2 & 29 & 1678 \\

\href{https://github.com/research-projects-all/master-spring-and-spring-boot}{GH-SB-Master} & 6,918 & 5 & 256 & 14 & 4 & 279 & 38 & 241 & 2 & 239 & 275 \\

\href{https://github.com/research-projects-all/spring-boot-3-jwt-security}{GH-SB-JWT} & 1,293 & 1 & 0 & 0 & 2 & 3 & 2 & 1 & 1 & 0 & 3 \\

\href{https://github.com/research-projects-all/spring-boot-3-spring-6-hibernate-for-beginners}{GH-SB-Hibernate} & 37,236 & 11 & 128 & 132 & 2 & 273 & 260 & 13 & 12 & 1 & 273 \\

\href{https://github.com/research-projects-all/spring-boot-vuejs}{GH-SB-Vue} & 1,590 & 3 & 262 & 1 & 5 & 271 & 34 & 237 & 5 & 232 & 271 \\

\href{https://github.com/research-projects-all/spring-boot-websocket-chat-demo}{GH-SB-WebSocket} & 328 & 2 & 5 & 0 & 1 & 8 & 7 & 1 & 1 & 0 & 8 \\

\href{https://github.com/research-projects-all/spring-data-examples}{GH-SB-Data} & 91,844 & 5 & 7 & 6 & 9 & 27 & 16 & 11 & 10 & 1 & 23 \\

\href{https://github.com/research-projects-all/spring-mvc-quickstart-archetype}{GH-SB-MVC} & 777 & 1 & 1 & 0 & 2 & 4 & 2 & 2 & 2 & 0 & 4 \\

\href{https://github.com/research-projects-all/spring-restbucks}{GH-SB-Restbucks} & 4,067 & 3 & 1 & 3 & 6 & 13 & 7 & 6 & 5 & 1 & 12 \\

\href{https://github.com/research-projects-all/tut-spring-security-and-angular-js}{GH-SB-SecurityAngular} & 10,119 & 2 & 497 & 1 & 3 & 503 & 27 & 476 & 5 & 471 & 502 \\

\midrule

\multicolumn{1}{@{}l}{\quad\emph{Total}} & 1,470,908 & 181 & 2,804 & 705 & 584 & 4,274 & 2,595 & 1,679 & 71 & 1,595 & 4,179 \\

\multicolumn{1}{@{}l}{\quad\emph{Mean}} & 63,952.5 & 7.9 & 121.9 & 30.7 & 25.4 & 185.8 & 112.8 & 73.0 & 3.1 & 69.3 & 181.7 \\

\multicolumn{1}{@{}l}{\quad\emph{Median}} & 1,322.0 & 2.0 & 3.0 & 3.0 & 2.0 & 11.0 & 7.0 & 5.0 & 2.0 & 0.0 & 9.0 \\

\multicolumn{1}{@{}l}{\quad\emph{SD}} & 236,890.6 & 19.9 & 251.4 & 84.9 & 107.2 & 410.6 & 353.2 & 164.7 & 3.1 & 164.8 & 405.2 \\

\multicolumn{1}{@{}l}{\quad\emph{Max}} & 1,140,914 & 93 & 925 & 386 & 517 & 1,688 & 1,657 & 614 & 12 & 612 & 1,678 \\

\multicolumn{1}{@{}l}{\quad\emph{Repos. $>0$}} & 23 & 17 & 16 & 16 & 21 & 22 & 18 & 19 & 19 & 11 & 22 \\

\midrule\midrule

\multicolumn{12}{@{}l}{\textbf{All repositories (60)}}\\

\multicolumn{1}{@{}l}{\quad\emph{Total}} & 1,713,207 & 350 & 4,206 & 1,055 & 998 & 6,609 & 3,750 & 2,859 & 422 & 2,331 & 6,097 \\

\multicolumn{1}{@{}l}{\quad\emph{Mean}} & 28,553.5 & 5.8 & 70.1 & 17.6 & 16.6 & 110.2 & 62.5 & 47.6 & 7.0 & 38.9 & 101.6 \\

\multicolumn{1}{@{}l}{\quad\emph{Median}} & 1,367.0 & 1.5 & 5.0 & 5.0 & 4.0 & 27.0 & 9.5 & 11.5 & 3.0 & 1.0 & 20.0 \\

\multicolumn{1}{@{}l}{\quad\emph{SD}} & 147,941.9 & 14.0 & 167.1 & 53.9 & 66.9 & 266.7 & 224.0 & 106.7 & 12.7 & 106.6 & 263.0 \\

\multicolumn{1}{@{}l}{\quad\emph{Max}} & 1,140,914 & 93 & 925 & 386 & 517 & 1,688 & 1,657 & 614 & 89 & 612 & 1,678 \\

\multicolumn{1}{@{}l}{\quad\emph{Repos. $>0$}} & 60 & 41 & 46 & 51 & 58 & 59 & 53 & 56 & 55 & 38 & 59 \\

\end{longtable}

\endgroup

\noindent{\tiny\textit{Note:} LOC is the nonblank, noncomment line count from \texttt{cloc}. CQL = CodeQL; Sem. = Semgrep PRO (Semgrep OSS excluded); Sonar = SonarCloud; Hot. = SonarCloud Security Hotspots; All = sum of the four configurations; Tut. = alerts in tutorial-related files (including tests); Other = alerts in non-tutorial files. Work. = alerts in \texttt{.github/workflows/}, including Security Hotspots; Dep. = alerts in dependency manifests and build metadata (e.g., \texttt{package-lock.json}, \texttt{requirements.txt}, \texttt{Pipfile.lock}, \texttt{pom.xml}, \texttt{build.gradle}); installed dependency code is classified as Other, but not Dep. Work. and Dep. are subsets of Other and are not added again to All. $\dagger$ marks the only repository with no alerts from all four configurations. CWE counts include rule-derived SonarCloud and Security Hotspot CWE associations. Repos. $>0$ counts repositories with a nonzero value in that column.\par}

\begin{tcolorbox}[resbox]

The four default configurations reported different numbers of alerts for the same tutorial repositories. Across repositories, alert counts were concentrated in a small number of repositories and did not consistently correspond to the number of lines of code.

\end{tcolorbox}

\subsection{Relevance of Alerts}

\label{ss:RelevanceOfAlerts}

\emph{Alert relevance} measures whether generated alerts target the code associated with a tutorial task. We classified alerts according to whether they targeted tutorial-related files or other repository files. Overall, 56.7\% of the alerts targeted tutorial-scoped files. Table~\ref{tab:tutorial-repositories} reports the counts of tutorial-scoped and non-tutorial alerts for each repository. The proportion of tutorial-scoped alerts ranges from 45.7\% for Semgrep to 80.6\% for Hotspots across the four configurations.

Most non-tutorial alerts targeted CI/CD workflow files and dependency manifests or build metadata (Table~\ref{tab:tutorial-repositories}). These files are part of the repository but do not necessarily belong to the application development task covered by the tutorial. For example, GitHub Security Actions can report security alerts in workflow files that configure CI/CD and security analysis. Such alerts may concern repository automation or dependency management rather than Django or Spring application code. These alerts also concern the maintenance and security posture of the system as a whole, not just the application. The results show that the tools report alerts for both tutorial code and supporting repository files.

\begin{tcolorbox}[resbox]

Alert scope differed across the four configurations. Hotspots had nearly twice the proportion of tutorial-scoped alerts as Semgrep. Across all configurations, nearly half of the alerts targeted files outside the tutorial scope, mainly dependency files, build files, and CI/CD workflows.

\end{tcolorbox}

\begin{table}[!htbp]

\centering

\small

\setlength{\tabcolsep}{3pt}

\caption{Distinct rules, their share of the reference rule pool, and the concentration of alerts in the five most frequent rules across all default-configuration alerts.}

\label{tab:rule-frequency}

\begin{tabular}{@{}lrrrrr@{}}

\toprule

Configuration &

\shortstack{Distinct\\rules} &

\shortstack{Reference\\rule pool} &

\shortstack{Rule ratio\\(\%)} &

\shortstack{Top-five\\alerts} &

\shortstack{Top-five\\(\%)} \\

\midrule

CodeQL & 39 & $41^{\S}$ & 95.1 & 209  & 59.7 \\

Semgrep  & 646 & $2{,}932^{\ddagger}$ & 22.0 & 1,062 & 25.2 \\

SonarCloud & 74 & $1{,}147^{\dagger}$ & 6.5 & 661  & 62.7 \\

Hotspots  & 34 & $1{,}147^{\dagger}$ & 3.0 & 681  & 68.2 \\

\bottomrule

\end{tabular}

\vspace{1mm}

\parbox{\columnwidth}{%

\scriptsize

$^{\dagger}$ SonarSource Python and Java rule catalogs(414 and 733 rules, respectively)~\cite{SonarSourcePythonRules2026,SonarSourceJavaRules2026}. $^{\ddagger}$ Semgrep Matching Rules under the active CI policy~\cite{semgrep1}. $^{\S}$ CodeQL Python query suite under the reported workflow configuration~\cite{codeql1}.}

\end{table}

\subsection{Rule Coverage}

\label{ss:RuleCoverage}

\emph{Rule coverage} describes the range of distinct security rules represented in the alerts that learners receive. We examined the number of distinct rules that produced alerts and the distribution of alerts across these rules for the four configurations. Table~\ref{tab:rule-frequency} reports the number of distinct rules, their proportion of the reference rule pool, and the proportion of alerts associated with the five most frequent rules for each configuration.

\paragraph{Rule Diversity and Concentration.}

The configurations differed in both the number of distinct rules triggered and the concentration of alerts among the most frequent rules. Semgrep triggered 646 distinct rules, compared with 39 for CodeQL, 74 for SonarCloud, and 34 for Hotspots. The five most frequent rules were 25.2\% of Semgrep alerts, compared with 59.7\% of CodeQL alerts, 62.7\% of SonarCloud alerts, and 68.2\% of Hotspots alerts. Thus, Semgrep alerts were distributed across a larger number of rules, whereas the five most frequent rules were more than half of the alerts in each of the other three configurations.

We also examined the rules that generated the highest number of alerts. For Semgrep, the most frequent rule was active debug code (373 alerts). The second most frequent rule was escaping in Django template translations (334 alerts). For CodeQL, the most frequent rule was GitHub Actions permissions (93 alerts). These examples show that the configurations differed not only in the number of distinct rules they triggered but also in the types of rules that generated the most alerts.

\paragraph{Common Weakness Enumerations (CWEs).}

We examined the \emph{CWE} identifiers associated with the alerts to identify the types of security weaknesses reported by each configuration and compare their frequencies. CodeQL and Semgrep provided CWE identifiers in their alert metadata. For SonarCloud and Hotspots, we obtained the CWE identifiers from the corresponding rules in the SonarSource rule database~\citep{sonar_rules}. Thus, the process of obtaining CWE information differed across the configurations. Table~\ref{tab:cwe-freq} reports the frequencies of the CWE identifiers for each configuration.

For Semgrep, cross-site scripting (CWE-79) caused 574 alerts, mostly in Django template files. Active debug code (CWE-489) was also frequently reported by both Semgrep and Hotspots, accounting for 373 and 311 alerts, respectively. CodeQL reported alerts related to access control and data flow, including URL redirection and workflow permissions (CWE-275). In addition, we observed CWE identifiers associated with code outside the primary web application development context. For example, the \texttt{GH-SB-JournalDev} repository contained Android-specific alerts related to improper component export (CWE-926).

\paragraph{OWASP Top 10:2025.}

\begin{figure}[!htbp]

\centering

\includegraphics[width=.95\linewidth]{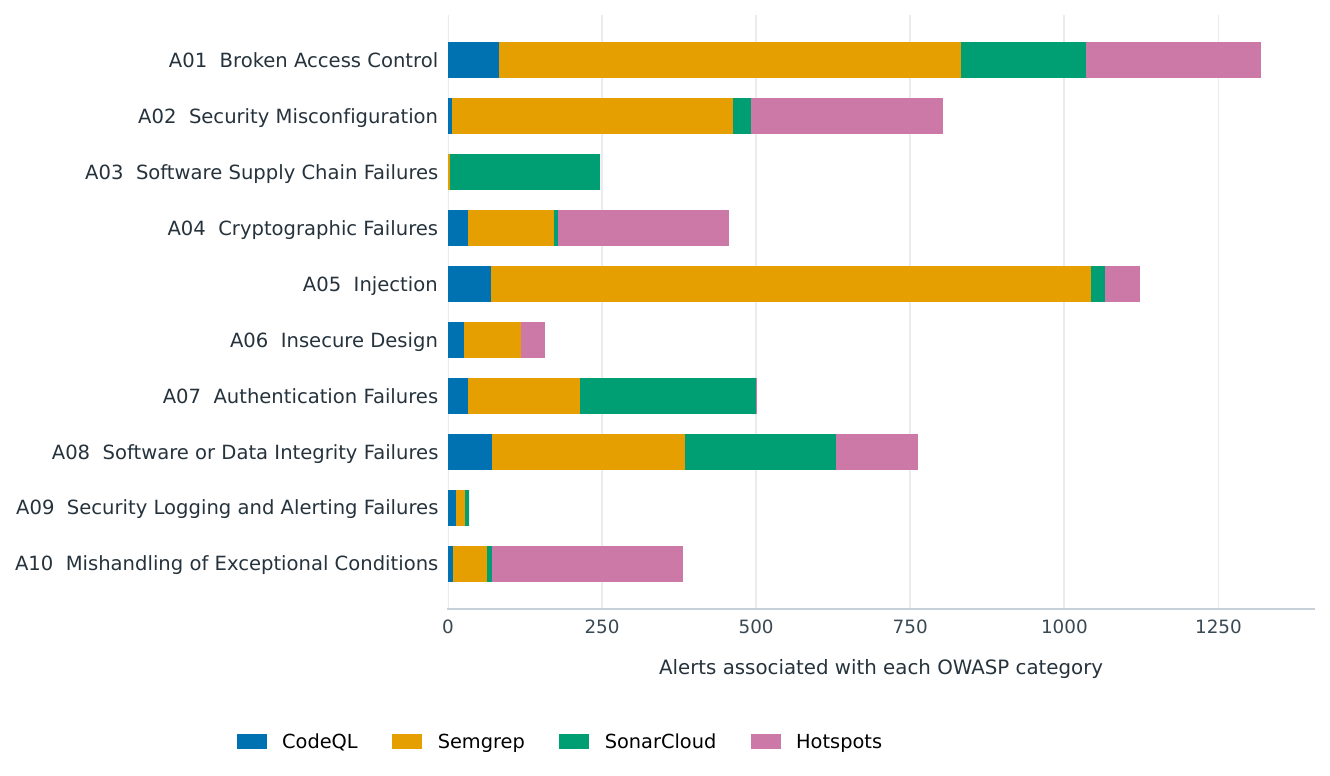}

\caption{OWASP Top 10:2025 category assignments across all four configurations.}

\label{fig:owasp25-distr}

\end{figure}

We mapped the CWE identifiers to the official CWE lists for the OWASP Top 10:2025 categories to examine which known web application security risks are represented in the alerts from each configuration. Figure~\ref{fig:owasp25-distr} shows the distribution of alerts across these categories.

For CodeQL and Semgrep, Injection (A05) and Broken Access Control (A01) were the major shares of mapped alerts. These alerts included cross-site scripting, cross-site request forgery, path traversal, and open redirects. SonarCloud showed a different distribution, with alerts related to hard-coded credentials, Android component export, and dependency version control. These alerts were associated with Authentication Failures (A07), Software or Data Integrity Failures (A08), and Software Supply Chain Failures (A03), respectively. Hotspots showed another distribution, with alerts associated with Mishandling of Exceptional Conditions (A10) and Cryptographic Failures (A04).

We also identified CWE identifiers that did not map to any OWASP Top 10:2025 category. For CodeQL, these included workflow permission alerts associated with CWE-275. For Semgrep, they included regular-expression complexity and prototype-pollution rules. Among Hotspots alerts, bidirectional control characters and regular-expression complexity are alerts without a mapped category. Thus, some alerts reported weaknesses not covered by the OWASP Top 10:2025 categories.

\begin{tcolorbox}[resbox]

Alerts differed in the number and types of security rules they triggered, with some configurations reporting alerts across many rules and others concentrating alerts around a few recurring rules. The alerts also differed in the security weaknesses and OWASP Top 10:2025 categories they represented.

\end{tcolorbox}

\begin{table*}[!htbp]

\centering\scriptsize

\caption{Most frequent CWE identifiers across the four default configurations. An alert can contribute to more than one row.}

\label{tab:cwe-freq}

\setlength{\tabcolsep}{5pt}

\begin{tabular}{@{}lrrrrr@{}}

\toprule

CWE & CodeQL & Semgrep & SonarCloud & Hotspots & Total \\

\midrule

CWE-79 & 57 & 574 & 11 & 52 & 694 \\

CWE-489 & 0 & 373 & 0 & 311 & 684 \\

CWE-1333 & 3 & 431 & 0 & 21 & 455 \\

CWE-798 & 0 & 83 & 272 & 0 & 355 \\

CWE-926 & 3 & 156 & 159 & 0 & 318 \\

CWE-215 & 0 & 0 & 0 & 311 & 311 \\

CWE-400 & 4 & 280 & 3 & 21 & 308 \\

CWE-200 & 1 & 102 & 1 & 203 & 307 \\

CWE-1321 & 0 & 293 & 0 & 0 & 293 \\

CWE-352 & 24 & 232 & 0 & 9 & 265 \\

CWE-319 & 28 & 28 & 0 & 203 & 259 \\

CWE-22 & 14 & 170 & 11 & 0 & 195 \\

CWE-922 & 0 & 0 & 158 & 0 & 158 \\

CWE-494 & 28 & 8 & 0 & 117 & 153 \\

CWE-829 & 30 & 1 & 0 & 117 & 148 \\

\bottomrule

\end{tabular}

\end{table*}

\begin{table*}[!htbp]

\centering

\small

\caption{Examples of alert messages from the tutorial repositories. We group the examples by the information present in each message. The first column identifies the alert and its reported location; the final column characterizes the message text.}

\label{tab:message-examples}

\setlength{\tabcolsep}{5pt}

\renewcommand{\arraystretch}{1.17}

\begin{tabularx}{\textwidth}{@{}>{\raggedright\arraybackslash}p{.255\textwidth}>{\raggedright\arraybackslash}X>{\raggedright\arraybackslash}p{.175\textwidth}@{}}

\toprule

\textbf{Alert and location} & \textbf{Exported message} & \textbf{Information in message} \\

\midrule
\multicolumn{3}{@{}l}{\emph{States a weakness or precaution without a named local fix}}\\
\midrule

\addlinespace[2pt]

\texttt{GH-DJ-Allauth}\\

\href{https://github.com/research-projects-all/django-allauth/security/code-scanning/25}{CodeQL; CWE-601}\\

\path{allauth/idp/oidc/views.py:176}

& \textit{``Untrusted URL redirection depends on a user-provided value.''}

& Identifies the risk; does not specify a code change. \\

\addlinespace[4pt]

\texttt{SB-BaeldungMVC}\\

\href{https://github.com/research-projects-all/bealdung-main-tutorial-4/security/code-scanning/9}{CodeQL; CWE-79}\\

\texttt{.../PathVariable}\allowbreak\texttt{Annotation}\allowbreak\texttt{Controller.java:84}

& \textit{``Cross-site scripting vulnerability due to a user-provided value.''}

& Names the weakness; does not specify a code change. \\

\addlinespace[4pt]

\texttt{GH-SB-JournalDev}\\

\href{https://sonarcloud.io/security_hotspots?id=research-projects-all_journaldev\&hotspots=AZ6oYjfVwChUXmJ39J9j}{Hotspots; \texttt{java:S4507}}\\

\path{.../AndroidCameraX/.../MainActivity.java:100}

& \textit{``Make sure this debug feature is deactivated before delivering the code in production.''}

& Recommends deactivation; does not name the setting. \\

\midrule
\multicolumn{3}{@{}l}{\emph{Names a code construct or a concrete change}}\\
\midrule

\addlinespace[2pt]

\texttt{GH-DJ-Allauth}\\

\href{https://github.com/research-projects-all/django-allauth/security/code-scanning/502}{Semgrep; CWE-352}\\

\path{.../templates/allauth/layouts/base.html:44}

& \textit{``Manually-created forms in django templates should specify a csrf\_token to prevent CSRF attacks.''}

& Names \texttt{csrf\_token} and where it belongs. \\

\addlinespace[4pt]

\texttt{SB-BaeldungMVC}\\

\href{https://github.com/research-projects-all/bealdung-main-tutorial-4/security/code-scanning/55}{Semgrep; CWE-352}\\

\texttt{.../ArticleViewer}\allowbreak\texttt{WithTwoSeparate}\allowbreak\texttt{Methods}\allowbreak\texttt{Controller.java:21}

& \textit{``Detected a method annotated with 'RequestMapping' that does not specify the HTTP method. \ldots{} To mitigate, add the 'method' field and specify the HTTP method (such as 'RequestMethod.POST').''}

& Names the annotation, missing field, and example value. \\

\addlinespace[4pt]

\texttt{GH-DJ-ECommerce}\\

\href{https://github.com/research-projects-all/eCommerce/security/code-scanning/76}{Semgrep; CWE-798}\\

\path{src/ecommerce/aws/conf.py:7}

& \textit{``AWS Access Key ID Value detected. This is a sensitive credential and should not be hardcoded here. Instead, read this value from an environment variable or keep it in a separate, private file.''}

& Identifies a credential and proposes two storage options. \\

\midrule
\multicolumn{3}{@{}l}{\emph{Describes a weakness in a third-party dependency}}\\
\midrule

\addlinespace[2pt]

\texttt{GH-DJ-Allauth}\\

\href{https://github.com/research-projects-all/django-allauth/security/code-scanning/537}{Semgrep; CWE-1333}\\

\path{package-lock.json:4749}

& \textit{``Affected versions of word-wrap are vulnerable to Inefficient Regular Expression Complexity. word-wrap is vulnerable to Regular Expression Denial of Service (ReDoS).''}

& Names the package and weakness, not a tutorial-code change. \\

\bottomrule

\end{tabularx}

\vspace{1mm}

\end{table*}

\subsection{Alert Descriptions}

\label{ss:AlertDescriptions}

\begin{figure}[!htbp]

\centering

\includegraphics[width=.79\linewidth]{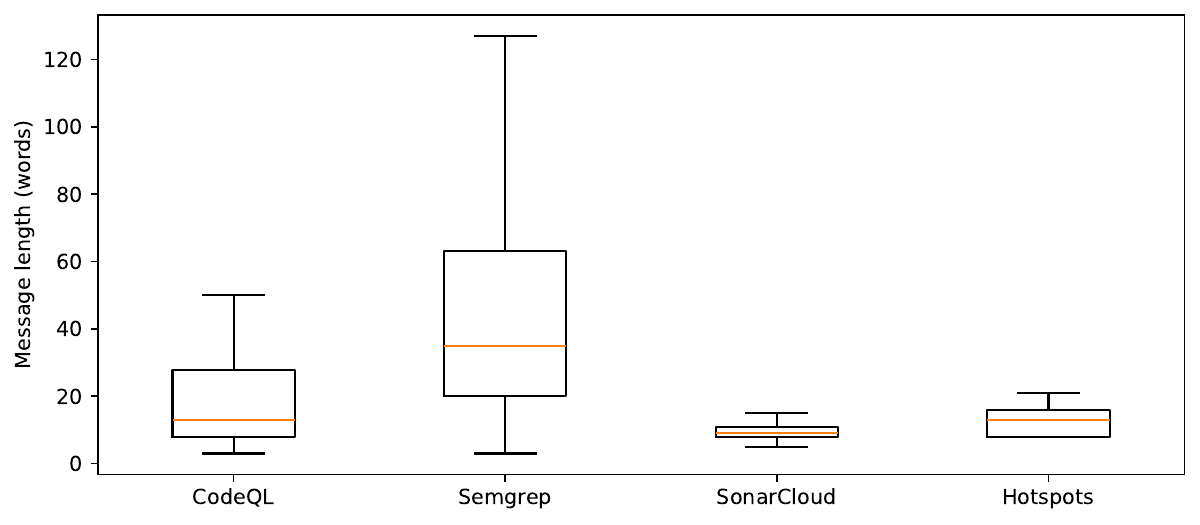}

\caption{Length of alert messages in words. Boxes show the interquartile range and median.}

\label{fig:message-dist}

\end{figure}

\begin{figure}[!htbp]

\centering

\includegraphics[width=.95\linewidth]{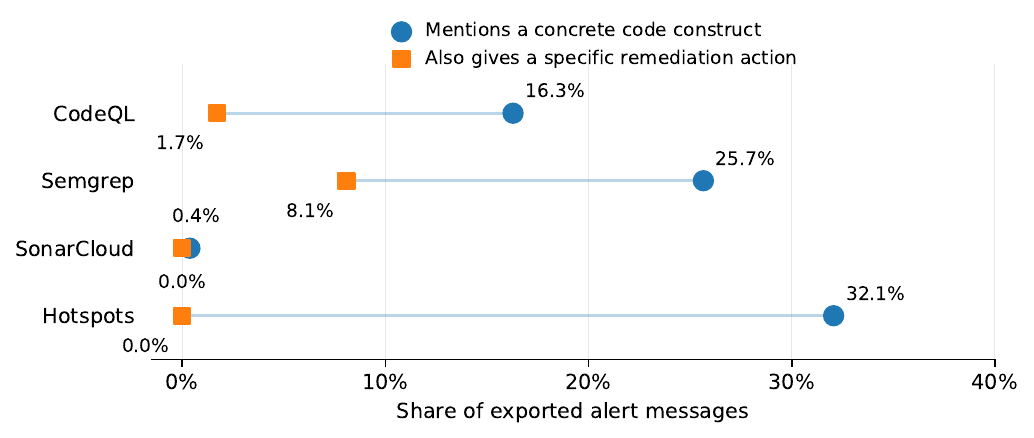}

\caption{Concrete-reference and remediation-specific rates in alert messages from the four default configurations. Percentages are calculated using all alerts from each configuration.}

\label{fig:rem-spec-distr}

\end{figure}

The alert description is the main text that learners read when they inspect an alert. Figure~\ref{fig:message-dist} shows the distribution of alert message lengths across the four configurations. Semgrep produced the longest messages, with a mean of 42.5 words, compared with 15.9 for CodeQL, 9.8 for SonarCloud, and 12.3 for Hotspots. We examined whether these descriptions identified the security issue, referred to concrete framework or code constructs, and provided specific remediation guidance. Across the configurations, concrete references and remediation-specific guidance appeared in a minority of alerts, as shown in Figure~\ref{fig:rem-spec-distr}.

Figure~\ref{fig:rem-spec-distr} shows differences in the information provided by the alert messages. Hotspots had the highest concrete-reference rate (32.1\%), followed by Semgrep (25.7\%), CodeQL (16.3\%), and SonarCloud (0.4\%). However, none of the Hotspots messages met both criteria for remediation specificity under our coding scheme. Semgrep had the highest remediation-specific rate (8.1\%), followed by CodeQL (1.7\%), while SonarCloud and Hotspots had rates of 0\%. Thus, mentioning a concrete code construct did not necessarily mean that a message also provided a specific remediation action.

Table~\ref{tab:message-examples} presents examples of alert messages from the different configurations. These examples illustrate the differences in concrete references and remediation guidance shown in Figure~\ref{fig:rem-spec-distr}. The CodeQL examples identify untrusted user input as a possible source of open redirects or cross-site scripting but do not specify how to validate the input or what code change to make. The Hotspots example recommends deactivating a debug feature before production but does not identify the relevant setting or explain how to deactivate it.

In contrast, several Semgrep messages refer to specific code constructs and suggest concrete changes. The Django CSRF message recommends including a CSRF token in manually created template forms. The Spring controller message identifies a missing HTTP method in \texttt{RequestMapping} and provides \texttt{RequestMethod.POST} as an example. The hardcoded credential message also suggests alternatives for storing the sensitive value, such as using an environment variable or a separate private file.

However, not all messages concern code directly related to the tutorial task. For example, a Semgrep alert in \texttt{package-lock.json} reports regular-expression complexity in the \texttt{word-wrap} dependency. The message identifies the affected package and describes the weakness, but it does not specify a change to the tutorial application code. This example illustrates how an alert can describe a security concern without providing remediation guidance connected to the code that the learner is developing. We also examined messages to characterize how the configurations presented their alert. Among the 998 Hotspots alerts, 540 (54.1\%) began with \emph{``Make sure''}, and 409 (41.0\%) contained the phrase \emph{``safe here''}. The most frequent Hotspots message, concerning debug features, occurred 278 times. Similarly, 385 of the 1,055 SonarCloud alerts (36.5\%) began with \emph{``Make sure''}. These recurring phrases indicate a review-oriented presentation style that asks developers to assess whether a security-related decision is appropriate. Such messages may identify a potential concern without specifying the framework-level change needed to address it.

\begin{tcolorbox}[resbox]

Alert messages varied in length, wording style, and remediation guidance. Concrete references appeared in 0.4-32.1\% of alerts, while remediation-specific content appeared in 0-8.1\% across configurations. Hotspots relied heavily on review-oriented wording: 54.1\% of messages began with \emph{``Make sure''}, and 41.0\% contained the phrase \emph{``safe here''}. These results show that naming a code construct does not necessarily provide learners with a concrete remediation step.

\end{tcolorbox}

\section{Discussion}
\label{sec:discussion}

Under default configurations, the GitHub Security Actions workflows we studied produced alerts outside tutorial code and triggered a broad range of security rules. Some alert messages also required prior security knowledge to interpret and act on. These results suggest that tutorial authors could adapt security alerts by selecting relevant rules and connecting alerts to the code learners study. They could also revise how messages explain security concerns.
\vspace{0.025em}

\noindent\textbf{Default security workflows may need adaptation for tutorial-based development.}
GitHub Actions let developers run security analysis within repository workflows~\citep{gitapi,action_required2026,unpacking_scanners2026}. However, this integration does not guarantee that alerts target the code that learners study. \citet{sast_for_review} found that SAST alerts did not always align with the code under review. \citet{li2025sast_eval} also showed that practitioners assess SAST tools against practical development needs. Our results also showed that running a security workflow did not necessarily produce alerts focused on tutorial code or provide concrete remediation guidance. Although GitHub Actions does not serve as an instructional tool, framework-specific references and actionable instructions in alerts could help developers connect reported alerts to code they can inspect and revise. Tutorial authors could review default alerts before before presenting them to learners. They could identify alerts relevant to the tutorial task and add framework-specific explanations when alerts lack concrete remediation steps. Default configurations can support broader repository-level security review for maintainers. Tutorial-scoped configurations could instead focus learners on security concerns relevant to a particular development task.

\medskip
\noindent\textbf{Security alert messages could offer direct guidance to learners.}
Our analysis identified recurring review-oriented wording in SonarCloud and Security Hotspots alerts. Such messages ask developers to assess whether a security-related decision is appropriate, but may not specify what they should inspect or change in the code. Prior work in software engineering recognizes alert presentation as an important aspect of development support~\citep{vandenaker2024formative}. For example, a SonarCloud Security Hotspot alert stated: \emph{``Make sure disabling Spring Security's CSRF protection is safe here.''} This message asks developers to assess a security decision without explaining how to assess it. Such an assessment requires knowledge of the application's authentication mechanisms and the security implications of disabling CSRF protection.~\citep{owasp-threat-modeling,spring-csrf-reference}. For learners, a more specific message could state: \emph{``Inspect the CSRF configuration in Spring Security, such as the \texttt{csrf} configuration within a \texttt{SecurityFilterChain}. Keep CSRF protection enabled for browser applications that use session-based authentication. Before disabling it for an API, check whether browsers automatically include authentication credentials in requests.''} This proposed wording identifies a framework-specific configuration that learners can locate. It also specifies conditions they can verify before making a security decision.

\medskip
\noindent\textbf{Tutorial authors could prioritize security rules according to learning objectives.}
Across the 60 repositories, Semgrep triggered 646 distinct rules. CodeQL triggered 39 rules and SonarCloud triggered 74. Hotspots triggered 34. A broad set of rules can expose learners to different security concerns, but unfamiliar concepts may require additional context and learning effort~\citep{congnitive_load,piskachev2023_can_configuration}. Our results suggest that tutorial authors could consider rule selection when adapting security alerts for tutorials. For example, authors could select rules based on the security concepts they want to prioritize in each lesson. GitHub reported an increase in vulnerabilities related to Broken Access Control~\citep{octoverse2025}, and OWASP identifies Broken Access Control as a leading web application security risk~\citep{owasp-broken-access-control}. A tutorial on web forms could focus on CSRF-related checks. Authors could introduce authorization-related checks when learners encounter access-control features. The default configurations we studied did not provide this instructional focus. Authors would need to examine the available rules and organize them according to their tutorial objectives. A broader rule set could remain available for repository-level security review. Such an approach might help learners revisit security concepts as their knowledge develops.

\medskip
\noindent\textbf{Alert scope could direct learners to the code they need to examine.}
We also found that security tools reported alerts outside tutorial code. For example, \texttt{GH-SB-SecurityAngular} had 503 alerts, but only 27 targeted tutorial-related files, while 471 were related to dependency files. Tutorial authors could map each lesson to its relevant files or code snippets and present alerts at those locations. For a Spring controller, a lesson-focused interface could point learners to the affected controller code and show the corresponding security alerts at that location. Alerts from frontend dependencies or workflow files could appear in a separate repository-level view. Unlike rule selection, this approach would not necessarily require changes to the security analysis itself. Instead, it could use the reported file paths and line ranges to locate alerts in code. A mapping between lessons and relevant code could then connect those alerts to the code learners currently study. Prior research identifies challenges developers face when interpreting alerts~\citep{christakis2016developers,usability-criteria}. Alerts outside the current lesson still deserve attention, but a lesson-focused presentation could help learners identify security concerns relevant to their current task.

\section{Conclusion}
\label{sec:conclusion}

Default GitHub Security Actions can be used to assess the security of tutorial code, but the alerts they produce do not consistently fit beginner-oriented learning conditions. We evaluated CodeQL, SonarCloud, SonarCloud Security Hotspots, and Semgrep on 60 Django and Spring tutorial repositories across different alert properties. Some \emph{repository} $\times$ \emph{tool} combinations produced manageable alert volumes, while others required intervention because they targeted non-tutorial artifacts such as workflow files and dependency metadata. Default tools triggered broad rule vocabularies, and their messages did not consistently connect alerts to concrete Django or Spring code changes. These alerts show a mismatch between default tool configuration assumptions and the needs of security beginners, who may not have broad code ownership, prior security knowledge, or the expertise needed to decide whether an alert applies. Our results can inform the design of security alerts for tutorials. In particular, they can inform how to scope analysis to tutorial-context code, curate rules around foundational web-security responsibilities, and include alert descriptions that have framework-specific remediation guidance. By adapting GitHub Security Actions to tutorial contexts, instructors, tutorial authors, and learning platforms can expose developers to secure coding practices earlier and reduce the chance that insecure patterns in tutorial code propagate into downstream projects and create vulnerabilities or attack vectors.

\section{Data Availability}
\label{sec:data-availability}

We provide the curated tutorial corpus, GitHub Actions workflows, raw alert data, annotations, analysis results, security-concept mappings, and statistical analysis code in our publicly available replication package~\citep{psp_replication_package}.

\bibliographystyle{spbasic}      

\bibliography{references-sast}

\end{document}